\documentclass[12pt]{article}
\pdfoutput=1

\usepackage{hyperref}

\usepackage[normalem]{ulem}
\usepackage{cite}
\usepackage{appendix}
\usepackage{epsfig}
\usepackage{amsmath}
\usepackage{amssymb}
\usepackage{bm}
\usepackage{hyperref}
\usepackage{slashed}
\newcommand{\be} {\begin{equation}}
\newcommand{\ee} {\end{equation}}
\newcommand{\ba} {\begin{eqnarray}}
\newcommand{\ea} {\end{eqnarray}}
\newcommand{\vareps}{\varepsilon}
\newcommand{\cpeps}{\epsilon^{\rm CP}}
\newcommand{\no} {\nonumber}
\newcommand{\cL} {\mathcal L}
\newcommand{\cA} {\mathcal A}
\newcommand{\cT} {\mathcal T}
\newcommand{\cd} {{\cdot }}

\newcommand{\GeV}{\text{GeV}}
\newcommand{\SU}{\text{SU}}
\newcommand{\U}{\text{U}}

\usepackage{color}
\definecolor{darkblue}{cmyk}{1,0.3,0,0.2}
\definecolor{violet}{cmyk}{0,1,0,0.2}
\hypersetup{colorlinks, bookmarksnumbered, citecolor=darkblue, linkcolor=darkblue, pdfstartview=FitH, urlcolor=darkblue, linktocpage}

\newcommand{\arXhref}[1]{\href{http://arxiv.org/abs/#1}{#1}}

\begin{document}

\begin{flushright}
 ZU-TH-43/14 \\
 December 2014
\end{flushright}

\thispagestyle{empty}

\bigskip

\begin{center}
\vspace{1.5cm}
    {\Large\bf  Pseudo-observables in Higgs decays} \\[1cm]
   {\bf Mart\'in Gonz\'alez-Alonso$^a$, Admir Greljo$^b$, \\[0.1cm]
   Gino Isidori$^{b,c}$, David Marzocca$^b$}    \\[0.5cm]
  {\em $(a)$  IPN de Lyon/CNRS, Universit\'e Lyon 1, Villeurbanne, France}  \\
  {\em $(b)$  Physik-Institut, Universit\"at Z\"urich, CH-8057 Z\"urich, Switzerland}  \\ 
  {\em $(c)$  INFN, Laboratori Nazionali di Frascati, I-00044 Frascati, Italy}  \\[1.0cm]
\end{center}

\centerline{\large\bf Abstract}
\begin{quote}
\indent
We define a set of pseudo-observables characterizing the properties of   Higgs decays in generic 
extensions of the Standard Model with no new particles below the Higgs mass.
The pseudo-observables  can be determined from experimental data,
providing a systematic generalization of the ``$\kappa$-framework" so far adopted 
by the LHC experiments. The  pseudo-observables are defined from on-shell decay amplitudes,  
allow for a systematic inclusion of higher-order QED and QCD corrections
and can be computed in any Effective Field Theory (EFT) approach to Higgs physics. 
We analyze the reduction of the number of  independent  pseudo-observables 
following from the hypotheses of lepton-universality,  CP invariance
, custodial symmetry, and linearly realized electroweak symmetry breaking. 
We outline the importance of kinematical studies of $h\to 4\ell$ decays for the extraction of such parameters 
and present their predictions in the linear EFT framework. 
\end{quote}
\vspace{5mm}

\newpage
\tableofcontents 

\newpage
\section{Introduction}

After the discovery phase~\cite{Higgs}, Higgs physics is entering the  era of precision measurements. 
Characterizing the properties of this particle with high precision, and possibly with the least theoretical 
bias, is of the utmost importance in order to investigate the nature of physics beyond the Standard Model (SM).

Several phenomenological analyses about the effective couplings of the Higgs boson to SM fields have 
appeared after its discovery in 2012 (see e.g.~Ref.~\cite{Azatov:2012bz}). These analyses were mainly based on the so-called
``$\kappa$-framework"~\cite{LHCHiggsCrossSectionWorkingGroup:2012nn}
 or ``signal-strength" results reported by ATLAS and CMS~\cite{ATLAS:2013sla,CMS:yva}:  the 
experimental determination of a single parameter, for each production or decay channel, characterizing the ratio between 
the observed rates and those expected within the SM. While this approach has been quite useful for a first characterization 
of the properties of the newly discovered particle, and it was appropriate given the low statistics so far available, it is 
insufficient in view of more precise studies, especially for channels with non-trivial kinematical distributions.
The purpose of the present paper is to provide a systematic 
generalization of the ``$\kappa$-framework" suitable for high-precision studies of on-shell Higgs decays. 

Motivated by present Higgs data, we work under the hypothesis that $h(125)$ is a spin zero particle. We also 
assume that there is no new particle with mass below (or around) $m_h \simeq 125 ~ \GeV$ able to provide significant kinematical 
distortions of the Higgs decays to SM particles.  In other words, we assume to be in a regime where the 
Effective Field Theory (EFT) approach to Higgs physics is applicable.  However, contrary to existing EFT studies,
we keep our analysis as general as possible, without  specifying many details about the underlying EFT. 
In particular, we do not specify if the $h(125)$ state is part of an $\SU(2)_L$ doublet (so-called linear EFT approach),
or if $h(125)$ is the mass eigenstate resulting from a more complicated symmetry breaking sector, allowing an
effective decoupling of $h$ from the Goldstone-boson components of the  $\SU(2)_L \times \U(1)_Y/ \U(1)_{\rm em}$ symmetry breaking
(so-called non-linear EFT approach). We  also do not impose global symmetry hypotheses such as 
lepton-universality,  CP invariance and custodial symmetry. Rather, we discuss how such hypotheses 
can be tested from Higgs data. The only key  assumption we make  is 
to  neglect terms in the decay amplitudes that receive non-vanishing tree-level contributions 
from local operators with dimension greater than six ($D>6$), as specified in detail in the following.

Under such general assumptions it is possible to define a limited set of pseudo-observables 
that can be directly determined from experimental data on Higgs physics and that encode all 
possible New Physics (NP) effects. These pseudo-observables are the natural generalization of the 
``$\kappa$-framework" so far adopted by the LHC experiments~\cite{LHCHiggsCrossSectionWorkingGroup:2012nn}, 
and an extension of the pseudo-observables 
employed to characterize NP effects in $Z$ physics at LEP~\cite{Bardin:1999gt,ALEPH:2005ab}.
The pseudo-observables are indeed defined at the amplitude level, allowing for a 
 systematic  inclusion of higher-order QED and QCD corrections: this leads to an accurate theoretical 
 description of Higgs decay amplitudes that recovers 
 the best up-to-date SM predictions in absence of NP effects.  The pseudo-observables thus determined from Higgs-physics data 
 can be computed in specific EFT approaches and, depending on the EFT employed, can possibly be
 correlated with non-Higgs physics observables for specific tests of the EFT approach.
 
 The paper is organized as follows: in Section~\ref{sec:general} we present a general discussion about Higgs decay
 amplitudes and pseudo-observables. In Section~\ref{sec:decomposition} we define the pseudo-observables 
characterizing Higgs decays mediated by electroweak gauge bosons. 
In Sections~\ref{sec:SM} and~\ref{sec:symmetry} we discuss the SM limit, the parameter counting, and the reduction 
of the number of  independent  pseudo-observables 
following from the hypotheses of lepton-universality,  CP invariance and custodial symmetry.
In Section~\ref{sec:h2e2m} we present a phenomenological analysis of the $h\to 2e2\mu$ channel,
focusing on the impact and the possible determination of the $h\to Z \bar\ell \ell$ contact terms. 
The results are summarized in the Conclusions. The Appendix~\ref{app:EFTmatch} contains the mapping between 
the pseudo-observables introduced in Section~\ref{sec:decomposition}
and the Wilson coefficients of $D=6$ operators in the linear EFT approach.
The Appendix~\ref{app:Custodial} contains an extended discussion about the constraints following from 
custodial symmetry. 

\section{General considerations}
 \label{sec:general}

Given the narrow width of the Higgs particle, the generic description of NP effects in 
processes involving one on-shell Higgs can be factorized into two parts: the production 
and the decay. In this work we concentrate on pseudo-observables characterizing 
the Higgs decay amplitudes, and we limit the attention to processes with at most four
particles in the final states (besides soft QED and QCD radiation).
To this purpose, we can distinguish two main categories: 
\begin{itemize}
\item[I.]helicity-violating decays into a pair of on-shell  fermions ($\bar b b $, $\tau^+\tau^-$,  \ldots).
\item[II.]helicity-conserving decays to four fermions, two fermions and a (hard) photon, and two photons ($4\ell$, $2\ell2\nu$, $\ell^+\ell^- \gamma$, $\gamma\gamma$, \ldots).
\end{itemize}
The definition of pseudo-observables for the first category is quite obvious, and will be presented 
at the end of this Section. The rest of the paper is devoted to the second category of decay
amplitudes, whose theoretical description in generic  EFT extensions of the SM is more involved. 

An early attempt to provide a general EFT-inspired description of $h\to 4\ell$ decay
amplitudes has been presented in Refs.~\cite{Isidori:2013cla,Isidori:2013cga}. Our work provides a generalization of the 
parametrization proposed there,
taking into account also the sub-leading effects of $Z\gamma$
and $\gamma\gamma$ intermediate states. We will also pay particular attention to a consistent 
separation of the  pseudo-observables accessible in Higgs  decays from those accessible via 
on-shell $Z$ or $W$ decays, defined in Sect.~\ref{sect:Zff}.
From this point of view,  our approach shares some similarities with the one
recently proposed in Ref.~\cite{primaries} (see also Ref.~\cite{Passarino:2012cb}).
However, we stress two conceptual 
differences with respect to Ref.~\cite{primaries}: 1)~our pseudo-observables are defined directly 
from the on-shell decay amplitudes and, as such, are unambiguously related to observable distributions;
2)~we make no assumptions about custodial symmetry and $\SU(2)_L$ properties of the $h$ 
particle. 
As anticipated, the only key hypothesis we employ is to neglect contributions to Higgs decay amplitudes 
corresponding to local interaction terms of $D>6$ after electroweak symmetry breaking.
More precisely, we employ the following simple power counting for each interaction term, based on its canonical dimension: $h$, gauge bosons, and 
derivatives (momenta) count as 1, while fermions count as $3/2$. With this counting we systematically neglect interaction terms with dimension $D>6$. 
This implies that our decomposition is able to accommodate all the effects generated,
at the tree-level, by the $D=6$ effective Lagrangian in the linear EFT framework (or the next-to-leading order terms in the expansion). 
Similarly, in the generic non-linear EFT framework, our decomposition is able to accommodate all 
the next-to-leading order terms in the expansion (disregarding single-Higgs interactions with $D\geq 7$). Even if the predicted size of each pseudo-observable 
varies depending on the specific EFT and its UV completion, the fact that interaction terms 
corresponding to higher-dimensional operators can be neglected is general (assuming no light NP).
Note also that while the decomposition is able to describe the effects generated at a given order in the EFT expansion, 
the pseudo-observables are defined  by the kinematical decomposition of the on-shell decay amplitudes and,
as such, they are well-defined independently of the EFT expansion. 

 Before proceeding, it is worth stressing that, in principle, there are two more categories of 
Higgs decay amplitudes affected by $D\leq6$ operators in a generic EFT approach:
\begin{itemize}
\item[III.]~helicity-violating amplitudes resulting from effective dipole interactions of the Higgs field to 
(light) fermions and electroweak gauge bosons.
\item[IV.]~four-quark final states resulting from the effective coupling of the Higgs to gluons. 
\end{itemize}
Even though there are no difficulties in including these in our formalism, we opt for not doing so
to keep our presentation more concise.

The first category is expected to be suppressed by light fermion masses in most realistic models 
and, independently of that, it does not interfere with the leading SM amplitudes in the limit of vanishing fermion masses. 
More precisely, we can neglect such amplitudes in the limit where we assume an exact 
$\U(1)_f$ symmetry acting on each of the light fermion species.\footnote{~In the lepton sector $f=e_L, e_R, \mu_L, \mu_R$, where the $\U(1)_{\ell_L}$ symmetries  $(\ell=e,\mu)$
act on the $\SU(2)_L$ doublets $(\ell_L, \nu^\ell_L)$.}
Note that such symmetry is a 
small sub-set of the full $\U(3)^5$ flavor symmetry often advocated in the EFT context: 
imposing such reduced symmetry group we can allow violations of lepton universality in the 
 $h\to 4\ell$ amplitudes ($\ell = e,\mu$), while consistently neglecting the helicity-violating dipole 
amplitudes and lepton flavor violating  interactions.

The second category is hardly accessible from the experimental point of view: the $hgg$ effective coupling 
is essential to determine the Higgs production cross section, but it cannot be identified via a well-measured 
Higgs partial decay width.

 \subsection{Pseudo-observables in $Z\to  f \bar f $ and $W\to  f  \bar f $ decays}
 \label{sect:Zff}
  
The SM charged and neutral current interactions are
\ba
\cL^{J}_{\rm SM} &=&    e  A_\mu J_{\rm em}^\mu+\frac{g}{c_w}  Z_\mu J_Z^\mu + \frac{g}{\sqrt{2}} 
\left( W^+_\mu J_+^\mu  +{\rm h.c.} \right)~,
\label{eq:one}
\ea
where 
\ba
\label{g}
J_{\rm em}^\mu &=& \sum_{f =f_L, f_R}   Q_f \bar{f } \gamma^\mu f ~, \no \\ 
J_Z^\mu &=& \sum_{f = f_L, f_R}  (T_3^f - Q_f s_w^2)  \bar f  \gamma^\mu f~,  \no \\ 
J_+^\mu &=& \sum_{\ell }    \bar \nu_{\ell L}  \gamma^\mu \ell_L +
 \sum_{u,d} 
  V_{ud}~ \bar u_L  \gamma^\mu d_L~,
\ea
$s_w=\sin\theta_W$, $c_w=\cos\theta_W$, $e =(4\pi \alpha_{\rm em})^{1/2}$ and $V_{ud}$ denote the elements of the Cabibbo-Kobayashi-Maskawa (CKM)
mixing matrix.

The effective interactions of the $Z$ and $W$ bosons to fermions are modified beyond the SM.
This effect can be taken into account by introducing appropriate effective couplings to describe 
the on-shell couplings of $Z$ and $W$   to fermions.
In particular,  we define the effective couplings $g_Z^f$,  $g_W^\ell$ and $g_W^{ud}$ as follows\footnote{In general, one could also write 
a right-handed coupling of $W$ boson to quarks; however, this is forbidden in the limit of unbroken $U(1)_{u_R}\times U(1)_{d_R}$
flavor symmetry.}
\ba \begin{split}
& \cA(Z(\vareps) \to f  \bar f ) =  i \sum_{f=f_L, f_R}   ~ g_Z^f  ~\vareps_\mu  ~ \bar f  \gamma^\mu f~,\\
& \cA(W^+(\vareps) \to      \ell^+ \nu  ) =  i   g_W^{\ell} ~ \vareps_\mu  ~ \bar \nu_{\ell L} \gamma^\mu \ell_L~, \qquad 
\cA(W^+(\vareps) \to u  \bar d  ) ~=  i  g_W^{ud}  ~\vareps_\mu  ~ \bar u_L  \gamma^\mu d_L~.
\label{eq:gZf}
\end{split}\ea
These effective couplings can be unambiguously determined from data using $Z$-pole observables ($Z$-boson partial decay widths, 
forward-backward or polarization asymmetries, together with the information on $m_Z$ from the $Z$ line shape), and on-shell $W$ 
decays.\footnote{In particular, LEP measurements at the $Z$ pole allow to set very precise constraints on the $Z$ couplings to each charged lepton, 
to neutrinos (summed over all possible light species), to the~$b$, $c$ and $u$ quarks~\cite{ALEPH:2005ab}, and a common coupling to the $s$ and $d$ quarks. 
Also the $W$ couplings to each lepton flavor, and 
a combination of the couplings to the light quarks can be constrained with high precision~\cite{Agashe:2014kda}.}
 As such, they are well-defined (basis-independent) pseudo-observables. In absence of rescattering effects,
the Hermiticity of the underlying effective Lagrangian implies that  
the $g_Z^f$ are real couplings,   while $g_W^{\ell}$ and  $g_W^{ud}$ can be complex.

These  pseudo-observables can be computed in any EFT. Within the SM, at the tree-level,
one finds
\ba
g_Z^{f, \rm SM} = \frac{g}{c_w} (T_3^f - Q_f s_w^2)~, \qquad
g_W^{\ell, \rm SM} = \frac{g}{\sqrt{2}}~, \qquad g_W^{ud, \rm SM} = \frac{g}{\sqrt{2}} V_{ud}~.
\ea

\subsection{Pseudo-observables in $h\to   f  \bar f$ decays}
\label{sect:hff}

In analogy to the effective couplings of $Z$ and $W$ bosons to fermions, 
for each fermion species we can introduce two real effective couplings ($y^f_{S,P}$)
defined by
\be 
\cA( h\to f \bar f) =   -\frac{i}{\sqrt{2}} \left[ ( y^f_S  +  i y^f_P)  \bar f_L  f_R +  ( y^f_S  - i y^f_P)  \bar f _R  f_L \right]~.
\ee
The ``dressing'' of this amplitude with  soft QED and QCD radiation is straightforward. 
The measurement of $\Gamma(h\to f\bar f)$ determines the combination $|y^f_S|^2 + |y^f_P|^2$,
while the $y^f_P/y^f_S$ ratio can be determined only if the lepton polarization 
is  experimentally accessible. If CP is conserved only one of the two effective couplings is allowed:
if $h$ is a CP-even state, then only $y^f_S$ is allowed. 

Within the SM, at the tree-level,
one finds
\ba
y^{f, \rm SM}_S =  \frac{ \sqrt{2} m_f}{v_F}~, \qquad
y^{f, \rm SM}_P =  0~, 
\ea
where $v_F = (\sqrt{2} G_F)^{-1/2} $,
 and  $G_F$ is the Fermi constant extracted from the muon decay. The 
 effective couplings $y^f_{S,P}$ provide an explicit breaking of the 
 $\U(1)_{f_L} \times  \U(1)_{f_R}$ flavor symmetry, which is not assumed to hold in the case 
 of third generation fermions. 

\section{Higgs decays mediated by electroweak gauge bosons} 
\label{sec:decomposition}

In this section we provide a unified decomposition of the Higgs decay amplitudes into 
four fermions ($h\to 4 f$), a fermion-anti fermion pair and one hard photon ($h\to  f\bar f \gamma $), 
and two photons ($h\to \gamma \gamma$).
The $h\to 4 f$ amplitudes are particularly interesting since they allow us to investigate 
the effective $hW^+W^-$ and $hZZ$ interaction terms, which cannot be probed on-shell. However,
in order to extract such information in a model-independent way, it is necessary to take into 
account also the possible additional contributions to $h\to 4 f$  due to contact terms 
and the effective couplings of the Higgs to photons. 

The purpose of our approach is to characterise, as precisely as possible, the three point function of the Higgs boson
 and two fermion currents, 
\be
	\langle 0 | {\cal T} \left\{ J_{f}^\mu(x),  J_{f^\prime}^\nu(y), h(0) \right\} | 0 \rangle~,
	\label{eq:ThreePointFunction}
\ee
where all the states are on-shell. This correlation-function  is probed by the experiments in $h \rightarrow 4f$ decays, 
but also in Higgs associated production ($pp\to h+W,Z$)  and in Higgs production via vector-boson fusion.
Extracting the kinematical structure of  Eq.~\eqref{eq:ThreePointFunction}  from data will allow us both
to determine the effective coupling of $h$ to all the SM gauge bosons, and also to investigate possible couplings of $h$ 
to new massive states. The former are associated to a well-defined double-pole structure in Eq.~\eqref{eq:ThreePointFunction},
while the latter can lead to local interactions with one or no poles. 

Within a generic EFT approach, the problem is simplified by the fact that a local interaction $h J_f^\mu J_{f^\prime}^\nu g_{\mu\nu}$ has canonical 
dimension $D = 7$.
As a result, as long as we neglect operators of $D> 6$, the correlation function in Eq.~\eqref{eq:ThreePointFunction}
is non-local at the electroweak scale, 
with at least one fermion pair generated by the propagation of one  electroweak gauge boson.
This allows us to decompose the  $h\to 4 f$ amplitude into a sum of neutral- and charged-current contributions,
according to the charge of  fermion current in Eq.~\eqref{eq:ThreePointFunction},
 and to expand around the physical poles produced by the propagation of the 
SM gauge bosons ($W, Z$ and $\gamma$). 
These two types of contributions are discussed separately in Sect.~\ref{sect:h4fneutral}
and Sect.~\ref{sect:h4fcharged}. The complete structure of a generic 
 $h\to 4 f$ amplitude  is presented in Sect.~\ref{sect:h4ffull}.
 
 \subsection{$h\to 4 f$ neutral currents}
 \label{sect:h4fneutral}
 
Let us consider the case of two different (light) fermion species: $h\to f\bar f + f^\prime \bar f^\prime$. 
As anticipated, we work in the limit of an exact $\U(1)_f \times \U(1)_{f^\prime}$ flavor symmetry.
In this limit, we can decompose the neutral-current contribution to the amplitude in the following way
\ba
&& \cA_{n.c.} \left[ h \to  f (p_1) \bar f (p_2) f^\prime  (p_3) \bar f^\prime (p_4) \right]  =  i \frac{2 m^2_Z}{ v_F}   \sum_{f = f_L, f_R}   \sum_{f^\prime = f^\prime_L, f^\prime_R} 
(\bar f  \gamma_\mu f) (\bar f^\prime  \gamma_\nu f^\prime) \cT^{\mu\nu} (q_1,q_2) \nonumber \\
&& 
\cT^{\mu\nu} (q_1, q_2) =  \left[ F^{f f^\prime}_1 (q_1^2, q_2^2) g^{\mu\nu} +  F^{f f^\prime}_3 (q_1^2, q_2^2)  \frac{ {q_1}\cd {q_2}~g^{\mu\nu} -{q_2}^\mu {q_1}^\nu }{m_Z^2}  
 +   F^{f f^\prime}_4 (q_1^2, q_2^2)  \frac{  \vareps^{\mu\nu\rho\sigma} q_{2\rho} q_{1\sigma}   }{m_Z^2}  \right], 
\no \\
 \label{eq:h4l1}
 \ea
where $q_1=p_1 +p_2$ and $q_2=p_3 +p_4$.  

From the assumption of no new light states in the EFT, and once again neglecting contributions from $D>6$ operators, we can decompose the form factors in  
full generality in the following way
\ba
F^{f f^\prime}_1 (q_1^2, q_2^2) &=& \kappa_{ZZ}  \frac{ g_Z^f  g_Z^{f^\prime}  }{P_Z(q_1^2) P_Z(q_2^2)}
  +  \frac{\epsilon_{Z f}}{m_Z^2}  \frac{ g_Z^{f^\prime}   }{  P_Z(q_2^2)} +    \frac{\epsilon_{Z f^\prime}}{m_Z^2}   \frac{ g_Z^f    }{  P_Z(q_1^2)}  +\Delta^{\rm SM}_{1} (q_1^2, q_2^2) ~,   \label{eq:h4lNeutrCurrEFT1}\\
F^{f f^\prime}_3 (q_1^2, q_2^2) &=&  \epsilon_{ZZ}  \frac{ g_Z^f  g_Z^{f^\prime}  }{P_Z(q_1^2) P_Z(q_2^2)}   +   \epsilon_{Z\gamma}  \left(      
 \frac{  e  Q_{f^\prime} g_Z^f   }{ q_2^2  P_Z( q_1^2) }   +  \frac{ e Q_f   g_Z^{f^\prime}   }{ q_1^2  P_Z( q_2^2) }  \right) +   \epsilon_{\gamma\gamma}   \frac{   e^2 Q_f Q_{f^\prime} }{ q_1^2 q_2^2  } \no \\
 && +\Delta^{\rm SM}_{3} (q_1^2, q_2^2),   \label{eq:h4l2b}
\\
F^{f f^\prime}_4 (q_1^2, q_2^2) &=&  \cpeps_{ZZ}  \frac{ g_Z^f  g_Z^{f^\prime}  }{P_Z(q_1^2) P_Z(q_2^2)}   +   \cpeps_{Z\gamma}  \left(      
 \frac{  e  Q_{f^\prime} g_Z^f   }{ q_2^2  P_Z( q_1^2) }   +  \frac{ e Q_f   g_Z^{f^\prime}   }{ q_1^2  P_Z( q_2^2) }  \right) +   \cpeps_{\gamma\gamma}   \frac{   e^2 Q_f Q_{f^\prime} }{ q_1^2 q_2^2  }~,~
 \label{eq:h4l2}
\ea
where  
 $g_Z^f$ are the effective couplings defined  in Eq.~(\ref{eq:gZf})  and 
 $P_Z(q^2) =  q^2 -m_Z^2 + i m_Z \Gamma_Z$.
Similarly to   $g_Z^{f}$, also $\kappa_{ZZ}$ and the $\epsilon_X$
are well-defined pseudo-observables that can be extracted from data and computed in any EFT.\footnote{~Here we generically denote by $\epsilon_X$ the 
parameters $\epsilon_{ZZ,Z\gamma, \gamma\gamma, Zf}$ and $\cpeps_{ZZ, Z\gamma, \gamma\gamma}$. } 
All the parameters but $\epsilon_{Z f}$ are flavor universal, i.e.~they do not 
depend on the fermion species.  In the limit where we neglect rescattering effects, both   $\kappa_{ZZ}$ and  $\epsilon_X$ are real.
The functions  $\Delta^{\rm SM}_{1,3} (q_1^2, q_2^2)$ encode non-local SM contributions generated beyond the 
tree level, which cannot be described in terms of $D \leq 6$ effective operators (see Sec.~\ref{sec:SM}).

Note that the  fact that the $g_Z^f$ are defined from on-shell $Z$ amplitudes is essential for $\kappa_X$ and $\epsilon_X$ to be 
well-defined physical quantities (independent of the choice of the EFT basis). 
Indeed, the decomposition  in Eqs.~(\ref{eq:h4l1}--\ref{eq:h4l2}) contains a set of $Z$-pole pseudo-observables
$\{ g_Z^f, m_Z, \Gamma_Z \}$, plus the low-energy 
input observables $\{G_F, \alpha_{\rm em}\}$, plus the set of Higgs-pole  pseudo-observables  $\{\kappa_{ZZ},\epsilon_X \}$.

 \subsection{$h\to 4f$ charged currents}
 \label{sect:h4fcharged}
 
Let's consider the $h\to \ell \bar \nu_{\ell}   \bar{\ell'} \nu_{\ell'}$ process.\footnote{~The analysis of a process involving quarks is equivalent, with the only difference that the 
$\epsilon_{Wf}$ coefficients are in this case non-diagonal matrices in flavor space, as the $g_{ud}^W$ effective couplings.} 
Employing the same assumptions used in the neutral current case, we can decompose the amplitude in the following way
\ba
&& \cA_{c.c.} \left[ h \to  \ell (p_1) \bar \nu_\ell (p_2)    \nu_{\ell'} (p_3)   \bar \ell^\prime  (p_4)  \right]  = i  \frac{2 m^2_W}{ v_F}   
(\bar \ell_L \gamma_\mu \nu_{\ell L} ) (\bar \nu_{\ell' L}  \gamma_\nu \ell^\prime_{L}) \cT^{\mu\nu} (q_1,q_2) \nonumber \\
&& 
\cT^{\mu\nu} (q_1, q_2) =  \left[ G^{\ell \ell'}_1 (q_1^2, q_2^2) g^{\mu\nu} +  G^{\ell \ell'}_3 (q_1^2, q_2^2)  \frac{ {q_1}\cd {q_2}~g^{\mu\nu} -{q_2}^\mu {q_1}^\nu }{m^2_W}  
 +   G^{\ell \ell'}_4 (q_1^2, q_2^2)  \frac{  \vareps^{\mu\nu\rho\sigma} q_{2\rho} q_{1\sigma}   }{m^2_W}  \right], 
\no \\
 \label{eq:h4lCharged}
\ea
where $q_1=p_1 +p_2$ and $q_2=p_3 +p_4$.

The EFT-inspired decomposition of the form factors is
\ba
G^{\ell \ell'}_1 (q_1^2, q_2^2) &=& \kappa_{WW}  \frac{ (g_W^{\ell})^*  g_W^{\ell'}  }{P_W(q_1^2) P_W(q_2^2)}
  +  \frac{(\epsilon_{W \ell})^*}{m_W^2}  \frac{ g_W^{\ell'}   }{  P_W(q_2^2)} + \frac{\epsilon_{W \ell'}}{m_W^2}  \frac{ (g_W^{\ell})^*   }{  P_W(q_1^2)} ~,   \\
G^{\ell \ell'}_3 (q_1^2, q_2^2) &=&  \epsilon_{WW}  \frac{ (g_W^{\ell})^* g_W^{\ell'}  }{P_W(q_1^2) P_W(q_2^2)} ~, \\
G^{\ell \ell'}_4 (q_1^2, q_2^2) &=&  \cpeps_{WW}  \frac{ (g_W^{\ell})^*  g_W^{\ell'}   }{P_W(q_1^2) P_W(q_2^2)} ~,
 \label{eq:h4lChargedEFT}
\ea
where $g_W^f$ are the effective couplings defined in Eq.~(\ref{eq:gZf}), and $P_W(q^2)$ is the $W$ propagator defined analogously to $P_Z(q^2)$.
In absence of rescattering effects,
the Hermiticity of the underlying effective Lagrangian implies that  
$\kappa_{WW}$,  $\epsilon_{WW}$ and $\cpeps_{WW}$ 
 are real couplings,   while $\epsilon_{W \ell}$ can be complex.

 \subsection{$h\to 4f$ complete decomposition}
 \label{sect:h4ffull}
  
The complete decomposition of a generic $h\to 4f$ amplitude is obtained combining neutral- and charged-current contributions depending on the 
nature of the fermions involved. For instance   $h \to 2 e 2\mu$ and $h \to \ell \bar \ell q \bar q$ decays are determined by a single neutral current amplitude,
while the case of two identical lepton pairs is obtained from  Eq.~(\ref{eq:h4l1}) taking into account the proper symmetrization of the amplitude:
\ba
\cA \left[ h \to  \ell (p_1) \bar \ell (p_2) \ell  (p_3) \bar \ell (p_4) \right] &=&     \cA_{n.c.} \left[ h \to  f (p_1) \bar f (p_2) f^\prime  (p_3) \bar f^\prime (p_4) \right]_{f=f^\prime=\ell} \no \\
& - &   \cA_{n.c.} \left[ h \to  f (p_1) \bar f (p_4) f^\prime  (p_3) \bar f^\prime (p_2) \right]_{f=f^\prime=\ell}~.
\ea
The $h \to  e^\pm  \mu^\mp \nu  \bar \nu $ decays receive contributions from a single charged-current amplitude,
while in the  $h \to  \ell \bar \ell  \nu \bar \nu $ case we have to sum charged and neutral-current contributions:
\ba
\cA \left[ h \to  \ell (p_1) \bar \ell (p_2) \nu  (p_3) \bar \nu (p_4) \right] &=&     \cA_{n.c.} \left[ h \to  \ell (p_1) \bar \ell (p_2)  \nu  (p_3) \bar \nu (p_4) \right]
\no \\
& - &   \cA_{c.c.} \left[ h \to   \ell (p_1) \bar \nu (p_4)  \nu (p_3)  \bar \ell  (p_2) \right]~.
\ea

\subsection{$h\to \gamma\gamma$ and $h\to f \bar f   \gamma $}
\label{sec:hgg}

The general form factor decomposition for these two channels is 
\ba
\cA \left[ h \to  \gamma (q, \epsilon)  \gamma(q^\prime, \epsilon^\prime) \right]  &=&   i \frac{2 }{ v_F}    \epsilon^\prime_\mu  \epsilon_\nu
 \left[    F^{\gamma\gamma}_3   (  g^{\mu\nu}~ {q}\cd{q^\prime} - {q}^\mu {q}^{\prime \nu}) 
 +   F^{\gamma\gamma}_4 \vareps^{\mu\nu\rho\sigma} q_{\rho} q^\prime_{\sigma}  \right], \qquad 
 \label{eq:h2gamma} \\
 \cA \left[ h \to  f (p_1) \bar f (p_2) \gamma (q, \epsilon) \right]  &=&   i \frac{2  }{ v_F}   \sum_{f = f_L, f_R}     (\bar f  \gamma_\mu f)  \epsilon_\nu 
\times \no \\
& \times & \!\!\!   \left[    F_3^{f\gamma} (p^2)  ( {p}\cd {q}~ g^{\mu\nu}-{q}^\mu {p}^\nu )
 +   F_4^{f\gamma} (p^2)     \vareps^{\mu\nu\rho\sigma} q_{\rho} p_{\sigma}  \right], 
 \label{eq:hZgamma} 
\ea
where $p=p_1+p_2$. After employing the EFT decomposition of the form factors, we do not need to  introduce 
additional parameters compared to the $h\to 4 f$ case:
\ba
&& F^{f\gamma}_3 (p^2) =    \epsilon_{Z\gamma}   \frac{   g_Z^f  }{ P_Z( p^2) }   +   \epsilon_{\gamma\gamma}   \frac{   e  Q_f  }{ p^2  } + \Delta_{3f\gamma}^{\rm SM}(p^2)~, \qquad 
 F^{\gamma \gamma}_3  =   \epsilon_{\gamma\gamma} ~, \qquad  \\
&& F^{f\gamma}_4 (p^2)  =    \cpeps_{Z\gamma}     \frac{   g_Z^f  }{ P_Z( p^2) }   +   \cpeps_{\gamma\gamma}   \frac{   e  Q_f   }{ p^2 }~, \qquad \qquad \qquad ~~
F^{\gamma \gamma}_4   =    \cpeps_{\gamma\gamma} ~. \qquad 
 \label{eq:h2l2}
\ea

\section{SM values}
\label{sec:SM}

Within the SM, at the tree level,
\be
\kappa_{ZZ}^{\rm SM-tree}  = 1~, \qquad \kappa_{WW}^{\rm SM-tree}  = 1~, \qquad \epsilon_{X}^{\rm SM-tree}  = 0~. 
\label{eq:SMtree}
\ee
One-loop electroweak corrections can be divided into two main categories: virtual QED corrections generated below 
the electroweak scale (after integrating out $W$, $Z$ and top-quark fields) and genuine virtual electroweak corrections
at the electroweak scale. The virtual QED corrections are sizable in various 
kinematical regions of $h\to 4 f$ and $h \to f \bar f\gamma$ decays and must be combined with the real radiation in order 
to obtain infrared safe observables. 
Their impact can be computed in a model-independent way for generic values of  $\kappa_X$ and $\epsilon_X$ 
(see Sect.~\ref{SMQED}).

The genuine electroweak corrections generate: i) small corrections to the tree-level
values of $\kappa_X$ and $\epsilon_X$ 
 in Eq. (\ref{eq:SMtree}); ii) small non-local contributions to the form factors;
iii) further tiny corrections that cannot be cast into the general decomposition in Eq.~(\ref{eq:h4l1}) and  (\ref{eq:h4lCharged}). 
These effects can be derived, in principle, by comparing our general decomposition
with the expression of the full SM next-to-leading order $h\to 4 f$ amplitude~\cite{Bredenstein:2006rh}.
As noted in Ref.~\cite{Chen:2014gka}, such corrections are very small (below the 1\% level compared to the tree-level
terms) and practically unobservable, except in a few notable kinematical points. In
particular, the only case where such corrections are relevant is for on-shell hard-photon 
amplitudes (given that they vanish at the tree-level within the SM) or almost on-shell photon-exchange
contributions in neutral-current amplitudes.

These effects are  taken into account by the SM one-loop $h\gamma\gamma$ and $hZ\gamma$
effective couplings~\cite{Bergstrom:1985hp} (see also Ref.~\cite{Djouadi:2005gi}),
that in our formalism read\footnote{~We introduce here the couplings $c_{Z\gamma} \simeq - 4.85$ and $c_{\gamma \gamma} \simeq - 6.49$ \cite{Grinstein:2013vsa}, defined from the effective Lagrangian 
\be
	\cL^{eff} = \frac{\alpha}{4\pi} \frac{h}{v} \left( \frac{c_{Z\gamma}}{s_w c_w} Z_{\mu\nu} F^{\mu\nu} + \frac{c_{\gamma \gamma}}{2} F_{\mu\nu} F^{\mu\nu} \right)~.
\ee
}
\be
	\epsilon^{\rm SM-1L}_{Z\gamma} = - \frac{\alpha}{4\pi s_w c_w} c_{Z\gamma}\simeq 6.7 \times 10^{-3}~,  \qquad 
	\epsilon^{\rm SM-1L}_{\gamma\gamma} = - \frac{\alpha}{4\pi} c_{\gamma \gamma}\simeq  3.8 \times 10^{-3}~,
	\label{eq:epsggSM}
\ee
and by the non-local terms $\Delta_3^{\rm SM}$ and $\Delta_{3f\gamma}^{\rm SM}$. 
The latter  can be decomposed as follows 
\be
\begin{split}
	\Delta_3^{\rm SM} (q_1^2, q_2^2) &= \Delta_{\gamma \gamma}^{\rm SM-1L}(q_1^2, q_2^2) \frac{e^2 Q_f Q_{f^\prime}}{q_1^2 q_2^2} + \Delta_{Z \gamma}^{\rm SM-1L}(q_1^2, q_2^2) \frac{e Q_{f^\prime} g_Z^f}{q_2^2 P_Z(q_1^2)} + \Delta_{Z \gamma}^{\rm SM-1L}(q_2^2, q_1^2) \frac{e Q_{f} g_Z^{f^\prime}}{q_1^2 P_Z(q_2^2)}~, \\
	\Delta_{3f\gamma}^{\rm SM} (p^2) &= \Delta_{\gamma \gamma}^{\rm SM-1L}(p^2, 0) \frac{e Q_f }{p^2} + \Delta_{Z \gamma}^{\rm SM-1L}(p^2, 0) \frac{g_Z^f}{P_Z(p^2)} ~
	\label{eq:4lSMcontr}
\end{split}
\ee
where the expressions of $\Delta_{Z \gamma,\gamma\gamma}^{\rm SM-1L}$ in the relevant kinematical region (i.e.~with at least one photon 
propagator close to be on-shell), are\footnote{~These approximated numerical expressions are precise at  the $1 \%$ level
for $q_1^2 \lesssim  (95 \GeV)^2$ in the case of  $\Delta_{\gamma \gamma}^{\rm SM-1L}$
and $ (30 \GeV)^2 \lesssim q_Z^2 \lesssim (120 \GeV)^2$ in the case of $\Delta_{Z \gamma}^{\rm SM-1L}$.}
\be\begin{split}
\Delta_{Z \gamma}^{\rm SM-1L}(q_Z^2, q_\gamma^2 \approx 0) &= \epsilon^{\rm SM-1L}_{Z\gamma} \left[  0.19  \frac{q_Z^2 - m_Z^2}{m_Z^2} +  0.05 \left( \frac{q_Z^2 - m_Z^2}{m_Z^2} \right)^2 +\ldots  \right], \\
\Delta_{\gamma \gamma}^{\rm SM-1L}(q_1^2, q_2^2 \approx 0) &= \epsilon^{\rm SM-1L}_{\gamma\gamma} \left[ 0.16 \frac{q_1^2}{m_Z^2} + 0.03 \left( \frac{q_1^2}{m_Z^2} \right)^2  +\ldots  \right]~.
\label{eq:SMformfactors}
\end{split}
\ee
Note that the $q^2$-dependent terms in Eq.~\eqref{eq:SMformfactors}  cancel one of the two propagators in $\Delta_3^{\rm SM} (q_1^2, q_2^2)$.
This implies that such terms can effectively be seen as contact interactions with a photon (of the type $h \gamma f \bar{f}$). 
However,  contrary to the contact terms appearing in $F^{f f^\prime}_1$, 
these contact terms receive contributions from EFT operators of $D\geq 7$ and therefore can be 
fixed to their SM values.

To make contact with the  $\kappa$-framework adopted by ATLAS and CMS~\cite{ATLAS:2013sla,CMS:yva},
we can trade the $\epsilon_{\gamma\gamma,\gamma Z}$ parameters for 
$\kappa_{\gamma\gamma, Z \gamma }$, defined by 
\be
\kappa_{\gamma\gamma ( Z \gamma) } = \frac{  \epsilon_{\gamma\gamma (Z \gamma) } }{ \epsilon^{\rm SM-1L}_{\gamma \gamma 
(Z\gamma )}}~,
\label{eq:kggzgObs}
\ee
such that  $\kappa_{\gamma\gamma, Z \gamma }^{\rm SM}  = 1$.

 
\section{Parameter counting and symmetry limits} 
\label{sec:symmetry}

We are now ready to identify the number of independent pseudo-observables necessary to describe various sets 
of Higgs decay amplitudes, under the main assumption that only terms arising at $D\leq6$ in a generic EFT expansion are kept.
We focus our attention on leptonic channels, which are more interesting from the experimental point of view.

The neutral current processes $h \rightarrow e^+ e^- \mu^+ \mu^-$, $h \rightarrow e^+ e^-  e^+ e^-  $ and $h \rightarrow \mu^+ \mu^- \mu^+ \mu^-$,
together with the  photon channels $h \to \gamma \gamma$ and $h \to \ell^+ \ell^- \gamma$,  can be described in terms of 11 real parameters:
\be
	\kappa_{ZZ},  \kappa_{Z\gamma},  \kappa_{\gamma \gamma},  \epsilon_{ZZ},  \epsilon_{ZZ}^{CP}, \epsilon_{Z\gamma}^{CP}, \epsilon_{\gamma \gamma}^{CP}, \epsilon_{Z e_L}, \epsilon_{Z e_R}, \epsilon_{Z \mu_L}, \epsilon_{Z \mu_R}
\ee
(of which only the subset  $\{ \kappa_{\gamma \gamma}, \kappa_{Z\gamma}, \epsilon_{\gamma \gamma}^{CP}, \epsilon_{Z\gamma}^{CP}, \}$
is necessary to describe $h \to \gamma \gamma$ and $h \to \ell^+ \ell^- \gamma$).
The charged-current process $h \rightarrow \bar{\nu}_e e \bar{\mu} \nu_\mu$ needs 7 further independent real parameters to be completely specified:
\be
	\kappa_{WW}, \epsilon_{WW}, \epsilon_{WW}^{CP}~ ({\rm real}) \quad  + \quad \epsilon_{W e_L},  \epsilon_{W \mu_L}~({\rm complex})~.
\ee
Finally, the mixed processes $h \rightarrow e^+ e^- \nu\bar{\nu}$ and $h \rightarrow \mu^+ \mu^- \nu\bar{\nu}$ can be described by a subset of the coefficients already introduced plus 2 further real contact interactions coefficients:
\be
	\epsilon_{Z \nu_{e}}, \epsilon_{Z \nu_{\mu}}~.
\ee
This brings the total number of (real) parameters to 20.
In the following subsections we introduce symmetry arguments which allow to reduce the number of free parameters while remaining, at the same time, as model-independent as possible.

\subsection{Flavor universality}

A first simple restriction in the number of parameters is obtained by assuming flavor universality (i.e.~enlarging the flavor symmetry to the full $\U(3)^5$ flavor group). 
In our setup this simply means assuming that the contact interactions coefficients are independent of the generations:
\be
	\epsilon_{Z e_L} = \epsilon_{Z \mu_L}~,\qquad
	\epsilon_{Z e_R} = \epsilon_{Z \mu_R}~,\qquad
	\epsilon_{Z \nu_{e}} = \epsilon_{Z \nu_{\mu}}~,\qquad
	\epsilon_{W e_L} = \epsilon_{W \mu_L}~.\qquad
\ee
Since the last coefficients are complex in general, these are five relations which allow to reduce the number of parameters to 15.
This assumption can be tested directly from data by comparing the extraction of the contact terms from 
$h \rightarrow 2e2\mu$, $h \rightarrow 4e$ and $h \rightarrow 4\mu$ modes (see e.g. Sect.~\ref{sec:ExpContTerms} and Fig.~\ref{fig:tot2}).

\subsection{CP conservation}

The assumption that CP is a good approximate symmetry of the BSM sector and that the Higgs is a CP-even state,
allows us to set to zero six independent (real) coefficients:
\be
	\epsilon_{ZZ}^{CP} = \epsilon_{Z\gamma}^{CP} = \epsilon_{\gamma\gamma}^{CP} = \epsilon_{WW}^{CP} = \text{Im} \epsilon_{W e_L} = \text{Im} \epsilon_{W \mu_L} = 0~.
\ee
Assuming, at the same time, flavor universality, the number of free real parameters reduces to 10.

\subsection{Custodial symmetry}

We now present the relations among the pseudo-observables introduced in Sect.~\ref{sec:decomposition} following from the assumption that the BSM sector is invariant under the custodial symmetry group $G = \SU(2)_L \times \SU(2)_R \times \U(1)_X$, spontaneously broken to the diagonal $H = \SU(2)_{L+R} \times \U(1)_X$. This symmetry is explicitly broken by the fact that only the subgroup $G_{\rm SM} = \SU(2)_L \times \U(1)_Y$ is gauged and by the fact that SM fermions are not in complete $G$ representations.\footnote{~The $\U(1)_X$ factor is needed only to assign the correct hypercharge $Y = T_{R}^3 + X$ to the SM fermions.} In the following we assume that these are the only two sources of breaking of custodial symmetry. 
In order to determine the structure of the contact interactions, we need to specify the embedding of the SM fermions into representations of $G$. 
Focussing on leptons, we consider two minimal cases: ($A$) $L_L^i \in ({\bf 2}, {\bf 1})_{-\frac{1}{2}}$, $e_R^i \in ({\bf 1}, {\bf 2})_{-\frac{1}{2}}$ and ($B$) $L_L^i \in ({\bf 2}, {\bf 2})_{-1}$, $e_R^i \in ({\bf 1}, {\bf 1})_{-1}$.\footnote{~Here and in the following we label by the index $i=1...3$ the three lepton generations and we denote by $L_L^i$ the lepton doublet $(e_L^i, \nu^i_L)^T$.}

\begin{table}[t]
\begin{center}
\begin{tabular}{|c||c|c|c|} \hline\hline
\rule{0pt}{1.2em}%
$h$ decay modes &  Maximal Symmetry & Flavor Non Univ. & CPV    \\
 \hline\hline
  $h \to \gamma\gamma,  2e \gamma, 2\mu\gamma$  &  $\kappa_{ZZ},  \kappa_{Z\gamma},  \kappa_{\gamma \gamma}$   & 
  \raisebox{-6pt}[0pt][0pt]{$\epsilon_{Z \mu_L}, \epsilon_{Z \mu_R}$}  &  
  \raisebox{-6pt}[0pt][0pt]{$\epsilon_{ZZ}^{CP}, \epsilon_{Z\gamma}^{CP}, \epsilon_{\gamma \gamma}^{CP}$}     
  \\
  $ 4e, 4\mu, 2e2\mu$ & $\epsilon_{ZZ}, \epsilon_{Z e_L}, \epsilon_{Z e_R}$ &   &   
  \\ \hline
   \raisebox{-10pt}[0pt][0pt]{ $h \to   2e 2\nu, 2\mu 2\nu, e\nu \mu \nu$} &  
  \raisebox{-3pt}[0pt][0pt]{ $\kappa_{WW}$ } & \raisebox{-4pt}[0pt][8pt]{ $\epsilon_{Z \nu_{\mu}}$, Re($\epsilon_{W \mu_L}$) } &  \raisebox{-4pt}[0pt][8pt]{$\epsilon_{WW}^{CP}$, Im($\epsilon_{W e_L}$)}
    \\  
   & $\epsilon_{WW}$, $\epsilon_{Z \nu_{e}}$, Re($\epsilon_{W e_L}$) 
     &   \multicolumn{2}{|c|}{  \raisebox{-2pt}[4pt][8pt]{
 Im($\epsilon_{W \mu_L}$)  }}  \\ \hline\hline
\raisebox{-4pt}[0pt][0pt]{ $h \to \gamma\gamma,  2e \gamma,  2\mu\gamma, 4e, 4\mu,$ } & & &  
  \\
\raisebox{-4pt}[0pt][0pt]{  $  2e2\mu, 2e 2\nu, 2\mu 2\nu, e\nu \mu \nu$ }
 &  \raisebox{8pt}[0pt][0pt]{  $\kappa_{ZZ},  \kappa_{Z\gamma},  \kappa_{\gamma \gamma}$}  &  $\epsilon_{Z \mu_L}, \epsilon_{Z \mu_R}$   
 &  \raisebox{-2pt}[0pt][0pt]{$\epsilon_{ZZ}^{CP}, \epsilon_{Z\gamma}^{CP}, \epsilon_{\gamma \gamma}^{CP}$}     
 \\ 
 & \raisebox{8pt}[0pt][0pt]{ $\epsilon_{ZZ}, \epsilon_{Z e_L}, \epsilon_{Z e_R}$}  &   &   
 \\ 
 \raisebox{4pt}[0pt][0pt]{  [with custodial symm.] } &  \raisebox{8pt}[0pt][0pt]{  Re($\epsilon_{W e_L}$) }  & &  \\
 \hline\hline
 \end{tabular}
\caption{\label{tab:POsumm} Summary of the pseudo-observables relevant to describe Higgs leptonic (and $\gamma\gamma$) decay modes.  In the second 
column (``Maximal Symmetry") we show the independent pseudo-observables needed for a given set of decay modes, 
assuming both CP invariance and flavor universality. The additional variables needed if we relax these symmetry hypotheses
are reported in the third and fourth columns. In the 
bottom row we show the independent pseudo-observables needed for a combined description of charged and neutral modes,
under the hypothesis of custodial symmetry.}
\end{center}
\end{table}

Under these assumptions, we derive the following custodial-symmetry relations among the pseudo-observables relevant to Higgs decays to four leptons 
\begin{eqnarray}
	\quad \epsilon_{WW} &=& c_w^2 \epsilon_{ZZ} + 2 c_w s_w \epsilon_{Z\gamma} + s_w^2 \epsilon_{\gamma \gamma}~, \label{eq:custVV}\\
	\quad \epsilon^{CP}_{WW} &=& c_w^2 \epsilon^{CP}_{ZZ} + 2 c_w s_w \epsilon^{CP}_{Z\gamma} + s_w^2 \epsilon^{CP}_{\gamma \gamma}~,\label{eq:custVVCP}\\
	\hspace{-0.5cm} \kappa_{WW} - \kappa_{ZZ}  &=& - \frac{2}{g} \left( \sqrt{2} \epsilon_{W e_L^i} + 2 c_w \epsilon_{Z e_L^i} \right)~,\label{eq:custkVV} \\
	\quad \epsilon_{W e_L^i} &=& \frac{c_w}{\sqrt{2}} (\epsilon_{Z \nu_L^i} - \epsilon_{Z e_L^i})~,\label{eq:custContL}\\
	 \epsilon_{Z e_R^i} &=&\epsilon_{Z \nu_L^i} + \epsilon_{Z e_L^i} \label{eq:custContR}~
	\qquad  {\rm [embedding~B~only]}~.
\end{eqnarray}
The first two relations have been derived first in Ref.~\cite{Contino:2013kra}; the complete derivation of all the relations can be found in Appendix~\ref{app:Custodial}. The first four are independent of the choice of the fermion embedding, while the last one is specific only for the embedding $B$.
We stress that  $\kappa_{WW} \not= \kappa_{ZZ}$ is consistent with custodial symmetry, given Eq.~\eqref{eq:custkVV}. The latter 
must be satisfied for any $i$ and implies 3 independent relations  in the case of 
flavor non universality. Assuming both flavor universality and CP invariance, the 
embedding-independent custodial symmetry relations lead to 3 independent constraints 
and allows us to decrease to 7 the number of free real parameters relevant to leptonic channels. 
The latter can be conveniently chosen as $\kappa_{\gamma\gamma}, \kappa_{Z\gamma},\kappa_{ZZ}, \epsilon_{ZZ},  \epsilon_{Z e_L}, \epsilon_{Z e_R} , 
{\rm Re}(\epsilon_{W e_L})$, as indicated in Table~\ref{tab:POsumm}.

\subsection{Linear vs. non-linear EFT}

In the SM the Higgs boson is part of an $\SU(2)_L$ doublet $H$ and the electroweak gauge symmetry is linearly realized. The \emph{linear} effective theory is built following this assumption: higher-dimensional operators are constructed in terms of the $H$ field~\cite{LinearEFT,Manohar:2006gz,Giudice:2007fh}.
This implies that the physical Higgs ($h$) appears in operators  contributing also to non-Higgs processes, and in particular to electroweak observables measured at LEP. In this context  it is thus possible to provide strong bounds on some Higgs observables using LEP data~\cite{Pomarol:2013zra,EFTHiggsFit,Contino:2013kra,primaries}.  A complete model-independent analysis of these (non-Higgs) constraints for the Higgs pseudo-observables  
is still missing, but we postpone it to a future work.

Assuming a linearly realized electroweak gauge symmetry provides also some relations among the Higgs pseudo-observables. These are due to an accidental custodial symmetry present in some of the $D=6$ operators. In particular, by matching our pseudo-observables with the coefficients of the $D=6$ operators, 
it turns out that the relations of 
Eqs.~\eqref{eq:custVV}, \eqref{eq:custVVCP} and \eqref{eq:custContL} are always exactly satisfied~\cite{Contino:2013kra}. 
This result implies that, independently of any symmetry assumption, the dynamical hypothesis of an underlying linear EFT 
reduces the number of relevant leptonic pseudo-observables from 20 to 14 (from 15 to 11 if flavor universality is further assumed).
In Appendix~\ref{app:EFTmatch} we derive these relations by an explicit matching with the operator basis of Ref.~\cite{Pomarol:2013zra}. Since the relations   derived involve only pseudo-observables, the result  is independent of the operator basis adopted.
The other two custodial symmetry relations, Eqs.~\eqref{eq:custkVV} and \eqref{eq:custContR}, are not satisfied in general in the linear EFT and turn out to be violated by non-vanishing coefficients of custodial symmetry violating operators (see App.~\ref{app:EFTmatchCust}).

\medskip

A more general approach to Higgs physics is  to build an EFT  allowing an
effective decoupling of $h$ from the Goldstone-boson components of the  $\SU(2)_L \times \U(1)_Y/ \U(1)_{\rm em}$ symmetry breaking.
In this case the electroweak symmetry is \emph{non-linearly} realized and the effective theory is built as a derivative expansion over the cutoff \cite{NonLinearEFT,Grinstein:2007iv,Alonso,Buchalla}. Given that the Higgs and the symmetry-breaking vev are independent, in this EFT it is not possible to connect 
electroweak observables (Higgs-less processes)  with Higgs observables. Moreover, since the Goldstone bosons 
are encoded in a dimensionless field, it is possible to write many more independent $D\leq 6$ operators than in the linear case. 
 It is easy to verify that in this context each pseudo-observabable of our parameterisation receives a 
non-vanishing tree-level contribution from an independent combination of effective operators.
In particular, it is  possible to 
  build custodially-violating operators~\cite{Alonso,Buchalla} that violate all the relations in Eqs.~\eqref{eq:custVV}--\eqref{eq:custContR}.

Even though electroweak precision tests and early Higgs data set strong constraints on the non-linear construction, favoring the linearly realized EFT, it is still early to draw a definite conclusion about this point. As  shown in Ref.~\cite{Isidori:2013cga}, $h\to 4\ell$ decays prove a very useful tool for settling this issue from data: 
if a violation of the electroweak bounds on the contact terms is measured, this will be a strong hint towards the non-linear realization. Given the above discussion, a similar conclusion could be derived in presence of a violation of the custodial symmetry relations in Eqs.~\eqref{eq:custVV}, \eqref{eq:custVVCP} and \eqref{eq:custContL}.

\begin{figure}
  \begin{center}
    \includegraphics[width=0.70\textwidth]{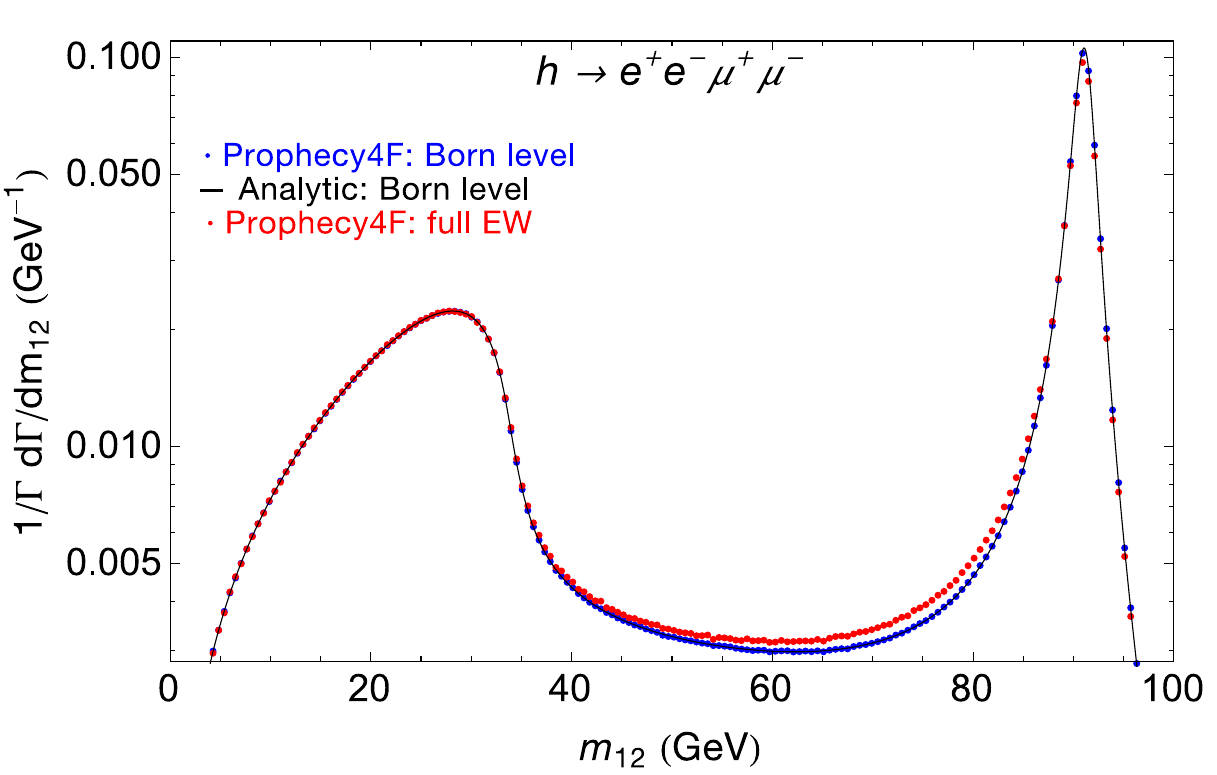}
  \end{center}
\caption{\label{fig:distrib} Normalized differential $h\to e^+e^- \mu^{+}\mu^{-}$ decay distribution in $m_{12}\equiv\sqrt{q_1^2}$ in the SM. Tree level predictions and full $\mathcal{O}(\alpha)$ electroweak corrections with Prophecy4F Monte Carlo generator~\cite{Bredenstein:2006rh} are shown with blue and red dots, respectively. The solid black line is obtained after integrating the analytic formula (Eq.~\ref{eq-diff-dis}) over $q_2^2$ for $\kappa_{ZZ}=1$
and $\epsilon_X=0$. }
\end{figure}

\section{Differential distributions for $h\to e^+e^- \mu^+\mu^-$}
\label{sec:h2e2m}

In this section we illustrate the importance of studying differential decay distributions for extracting the 
pseudo-observables defined in Section~\ref{sec:general}. We concentrate on the Higgs boson decay to pairs of muons and electrons,
which is particularly clean and possesses non-trivial kinematics. As a first step, we calculate the modification of the total decay rate to $e^+e^- \mu^+\mu^-$ keeping  only terms linear in $\epsilon_X$ and $\delta\kappa_{ZZ} \equiv \kappa_{ZZ} - 1$. We find
\be
\frac{\Gamma_{e^+e^- \mu^+\mu^-}}{\Gamma^{SM}_{e^+e^- \mu^+\mu^-}}=1+2\delta\kappa_{ZZ}-2.5\epsilon_{Ze_{R}}+2.9\epsilon_{Ze_{L}}-2.5\epsilon_{Z\mu_{R}}+2.9\epsilon_{Z\mu_{L}} + 0.5 \epsilon_{ZZ} - 0.9\epsilon_{Z\gamma} + 0.01 \epsilon_{\gamma\gamma}~.
\label{eq:Gtot}
\ee
Obviously, the measurement of the total rate is not enough to  extract the pseudo-observables and one should exploit the full kinematics of the process.

\subsection{Analytic invariant mass distributions}

In the following  we derive fully analytic expressions for the double differential decay distribution in each lepton pair's invariant mass.
Starting with Eq.~\eqref{eq:h4l1}, we calculate the matrix element squared and summed over the
final lepton spins,
\be
\sum_{\textrm{s}}\mathcal{A}\mathcal{A}^{*}=\left(\frac{2m_{Z}^{2}}{v_{F}}\right)^{2}\sum_{f,f'}tr(\slashed{p}_{1}\gamma_{\mu}P^{f}\slashed{p}_{2}\gamma_{\mu_{1}})
tr(\slashed{p}_{3}\gamma_{\nu}P^{f'}\slashed{p}_{4}\gamma_{\nu_{1}})\mathcal{T}_{ff'}^{\mu\nu}(q_{1},q_{2})\mathcal{T}_{ff'}^{\mu_{1}\nu_{1}*}(q_{1},q_{2}),
\ee
where $q_1 = p_1 + p_2$, $q_2 = p_3 + p_4$, $f = e_L,e_R$, $f^\prime = \mu_L, \mu_R$ and $P^{f}$ and $P^{f'}$ are the corresponding chirality
projection operators.
After integrating over the angular variables, we obtain an analytic formula for 
the double differential decay distribution in $q_1^2$ and $q_2^2$,
\be
\frac{d\Gamma}{dq_{1}^{2}dq_{2}^{2}}=\Pi_{4l}\int d\Omega\sum_{\textrm{s}}\mathcal{A}\mathcal{A}^{*},
\ee
where $\Pi_{4l}$ is the final state four body phase space factor.

The CP-conserving part of the double differential distribution can then be decomposed as 
\be
\frac{d\Gamma}{dq_{1}^{2}dq_{2}^{2}}= \frac{d\Gamma^{11}}{dq_{1}^{2}dq_{2}^{2}} + \frac{d\Gamma^{13}}{dq_{1}^{2}dq_{2}^{2}} +
\frac{d\Gamma^{33}}{dq_{1}^{2}dq_{2}^{2}}~,
\label{eq-diff-all}
\ee
where 
\ba
\frac{d\Gamma^{11}}{dq_{1}^{2}dq_{2}^{2}} &=& \frac{\lambda_{p}}{2^{10}(2\pi)^{7}m_{h}}\left(\frac{2m_{Z}^{2}}{v_{F}}\right)^{2}\frac{128\pi^{2}}{9}q_{1}^{2}q_{2}^{2}\frac{3+2\beta_{1}\beta_{2}-2(\beta_{1}^{2}+\beta_{2}^{2})+3\beta_{1}^{2}\beta_{2}^{2}}{(1-\beta_{1}^{2})(1-\beta_{2}^{2})}\sum_{f,f'} \left| F_{1}^{ff'} \right|^2~,
\no \\  \label{eq-diff-dis} \\  
\frac{d\Gamma^{13}}{dq_{1}^{2}dq_{2}^{2}} &=& \frac{\lambda_{p}}{2^{10}(2\pi)^{7}m_{h}}\left(\frac{2m_{Z}}{v_{F}}\right)^{2}\frac{128\pi^{2}}{3}(q_{1}^{2}q_{2}^{2})^{3/2}\frac{1+\beta_1 \beta_2}{\sqrt{(1-\beta_1^2)(1-\beta_2^2)}}2\sum_{f,f'} \text{Re}\left[F_{1}^{ff'}F_{3}^{ff'*} \right]~,
\no \\ \label{eq-diff-dis-hVV} \\
\frac{d\Gamma^{33}}{dq_{1}^{2}dq_{2}^{2}}&=&\frac{\lambda_{p}}{2^{10}(2\pi)^{7}m_{h}}\left(\frac{2}{v_{F}}\right)^{2}\frac{128\pi^{2}}{9}(q_{1}^{2}q_{2}^{2})^{2} \frac{3+4\beta_{1}\beta_{2}-(\beta_{1}^{2}+\beta_{2}^{2})+3\beta_{1}^{2}\beta_{2}^{2}}{(1-\beta_{1}^{2})(1-\beta_{2}^{2})}\sum_{f,f'} \left| F_{3}^{ff'}  \right|^2~,
\no \\
\label{eq-diff-dis-tens}
\ea
and
\be
\lambda_{p}=\sqrt{1+\left(\frac{q_{1}^{2}-q_{2}^{2}}{m_{h}^{2}}\right)^{2}-2\frac{q_{1}^{2}+q_{2}^{2}}{m_{h}^{2}}}~,
\qquad 
\beta_{1(2)}=\sqrt{1-\frac{4q_{1(2)}^{2}m_{h}^{2}}{(q_{1(2)}^{2}-q_{2(1)}^{2}+m_{h}^{2})^{2}}}~.
\ee

Using the explicit expressions of  $F_{1}^{ff'}$ and  $F_{3}^{ff'}$ in  Eqs.~\eqref{eq:h4lNeutrCurrEFT1}--\eqref{eq:h4l2b}
leads to a second order polynomial in $\kappa_X$ and $\epsilon_X$ for each value of $q_1^2$ and $q_2^2$.
Under the hypothesis of an underlying EFT,  only the interference terms of NP with the SM amplitude are expected to be relevant in a large fraction 
of the phase pace. If this were not the case, the approximation of neglecting terms in the amplitudes corresponding to higher-dimensional operators would 
not be justified. However, we stress that our parameterisation is equivalent  to a kinematical expansion of the amplitude around the physical poles 
of the SM gauge bosons. Sufficiently close to such poles it is possible to disregard the non-pole enhanced terms simply by kinematical arguments 
and organise a different power-counting for the momentum expansion of the rate. For instance,  requiring the $\mu^+\mu^-$ pair to be close the $Z$ peak allows us to 
consistently keep quadratic terms in $\delta \kappa_{ZZ}$, $\epsilon_{ZZ}$, $\epsilon_{Z e_{L,R}}$, and $\epsilon_{Z \gamma}$, while 
neglecting all other quadratic terms as well as the effect of $D>6$ interaction terms.
This fact could allow, in the future, to perform consistency checks about the validity of the EFT expansion. 

The tensor structure associated with $F_4$ of Eq.~\eqref{eq:h4l2} does
not interfere with the SM in the double differential distribution in $q_1^2$ and $q_2^2$.
In the rest of the paper we focus on the effects due to $\delta \kappa_{ZZ}$ and the contact terms $\epsilon_{Z f}$, 
leaving a more detailed phenomenological 
study of the other coefficients, which should involve also the analysis of angular 
distributions, to a future work. 

Further kinematical studies on the $h\to 4\ell$ modes can be found 
in Ref.~\cite{h4lep,Khachatryan:2014kca}. CMS performed a comprehensive study of  $h\to 4\ell$ decays with present data, in the context of $hVV$ anomalous couplings \cite{Khachatryan:2014kca}. The main differences of the latter approach with respect to our proposal is the fact that we 
consider as final states the on-shell leptons, and we do not assume the effective interaction of these leptons to the Higgs and other SM
fields to be necessarily mediated by the SM gauge bosons. This leads to a more general decomposition of the $h\to 4\ell$ amplitude.

\subsection{Higher-order SM corrections}
\label{SMQED}

We have validated the analytic formula for the tree level SM prediction with the Prophecy4F 
Monte Carlo generator~\cite{Bredenstein:2006rh}. In  Fig.~\ref{fig:distrib}, we present the
normalized differential distribution in $m_{12}\equiv \sqrt{q_1^2}$.
The solid black line corresponds to the results obtained after integrating Eq.~\eqref{eq-diff-dis} over 
$q_2^2$ for $\kappa_{ZZ}=1$ and $\epsilon_X=0$, while lowest order Prophecy4F predictions are 
shown with blue dots. The two predictions are in perfect agreement. Full $\mathcal{O}(\alpha)$ electroweak corrections obtained with Prophecy4F
are shown with red dots. Prophecy4F results are obtained after generating 
$10^8$ weighted events using the dipole subtraction formalism for photon radiation 
and switching on the photon recombination which insures sufficient 
inclusiveness~\cite{Bredenstein:2006rh}. More specifically, photons and leptons are 
recombined if their invariant mass is less then $5$~GeV. We impose no cuts on the decay products.

As shown in Fig.~\ref{fig:distrib}, the next-to-leading order electroweak corrections lead to a significant 
(up to 10\%) deviation from the tree-level result in the region below the $Z$ peak. This effect can well be understood 
in terms of photon emission from the charged leptons legs: radiative events where $m_{\ell^+\ell^-\gamma} \approx m_Z$
(close-to-on-shell $Z$-boson events) are enhanced by the  $Z$ pole but are 
reconstructed in the  $m_{\ell^+\ell^-}$ distribution as off-peak events 
($m_{\ell^+\ell^-} < m_Z$) providing a sizable distortion to the region below the $Z$ peak.
This effect can be corrected in general terms (for general values of the pseudo-observables) 
convoluting the non-radiative distribution with the $O(\alpha)$ radiation function describing the 
probability of emitting a photon (similarly to the initial-state radiation in $e^+e^- \to \mu^+\mu^-$ close to the $Z$ peak,
see e.g. Ref.~\cite{Nicrosini:1986sm}). We have explicitly checked that the inclusion of these corrections 
leads to a qualitatively good agreement of our results (in the SM limit) with the full $\mathcal{O}(\alpha)$ 
electroweak corrections obtained with Prophecy4F. The detailed implementation of these corrections,
which depends on the specific infrared cuts implemented in Prophecy4F, is beyond the scope of the 
present paper and will be discussed elsewhere.

\begin{figure}
  \centering
  \begin{tabular}{cc}
    \includegraphics[width=0.49\textwidth]{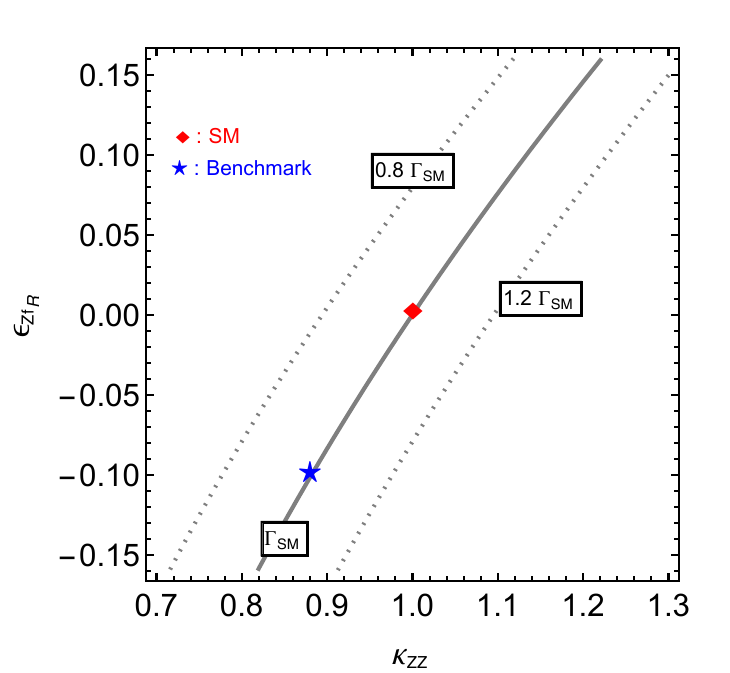} &
    \includegraphics[width=0.455\textwidth]{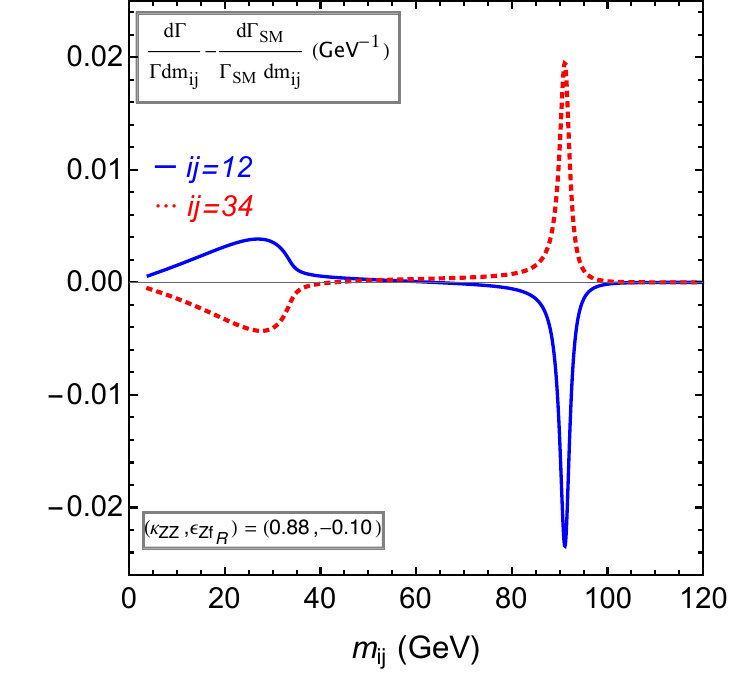} \\
            (a) & (b) \\
  \end{tabular}
   \includegraphics[width=0.54\textwidth]{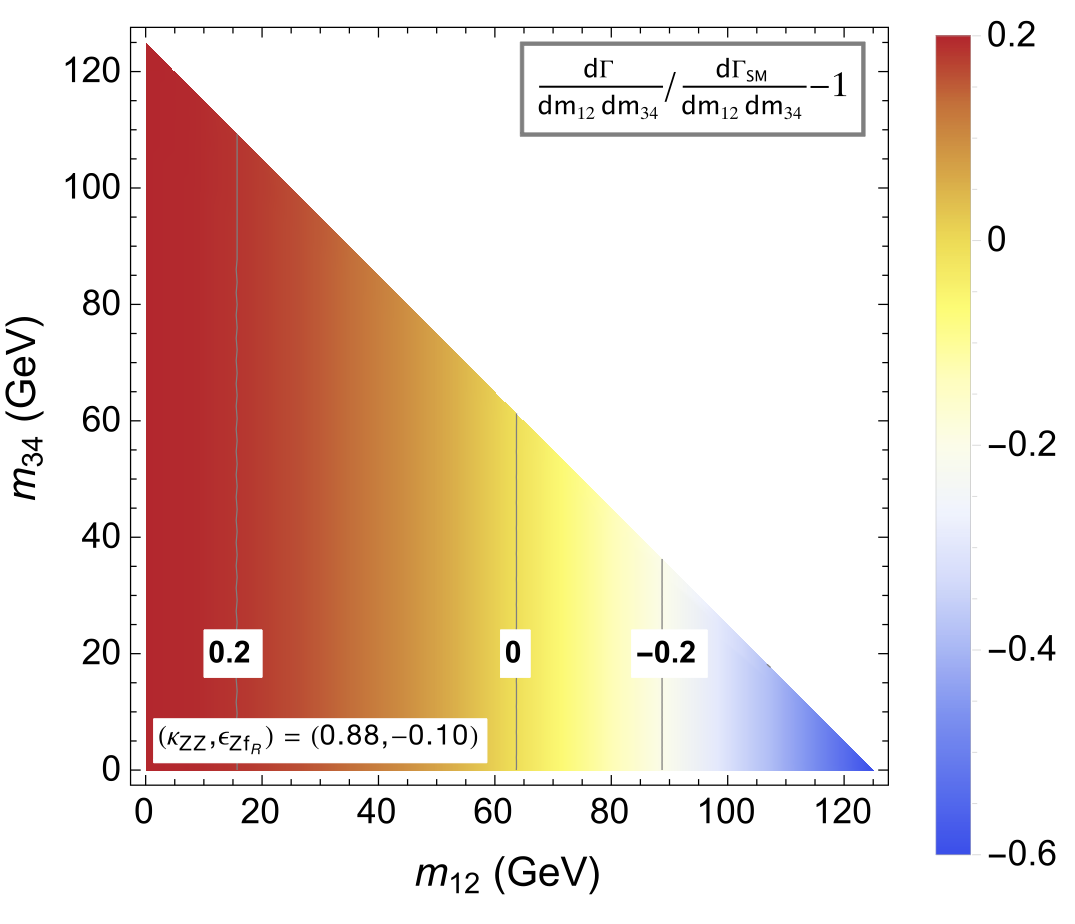} \\ 
   		(c)
  \caption{\label{fig:tot} (a) Total decay rate for $h\to e^+e^- \mu^+\mu^-$ as a function
  of $\kappa_{ZZ}$ and $\epsilon_{Zf_R}$ in units of the SM-predicted rate. (b) Deviations in the normalized single differential distributions in 
  $m_{12}\equiv \sqrt{q_1^2}$ (solid-blue line) and $m_{34}\equiv \sqrt{q_2^2}$ (dashed-red line) from the 
  SM expectations for the benchmark point $(\kappa_{ZZ},\epsilon_{Zf_R})=(0.88,-0.10)$. (c) Ratio of the double differential distribution with the SM prediction for the same benchmark point. No cuts are applied on the Higgs decay products.}
\end{figure}

\begin{figure}
  \centering
  \begin{tabular}{cc}
    \includegraphics[width=0.49\textwidth]{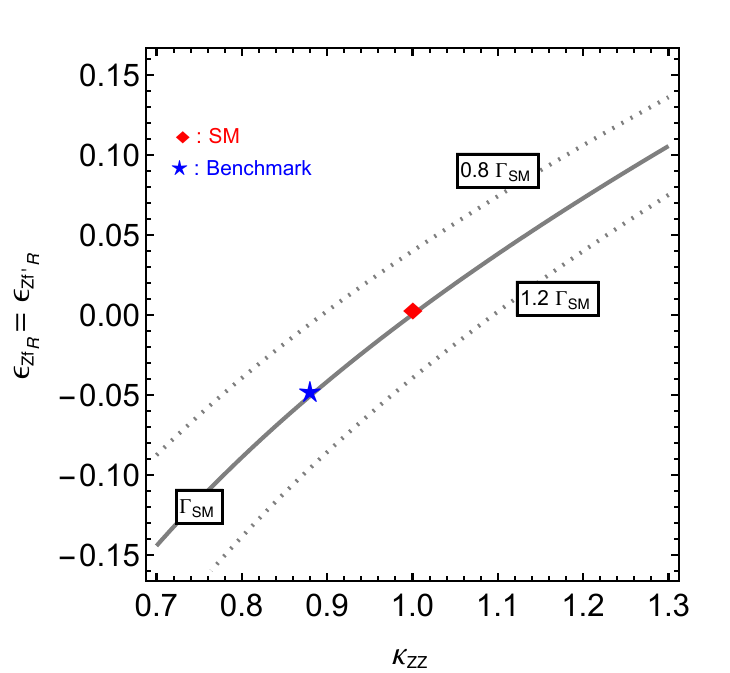} &
    \includegraphics[width=0.47\textwidth]{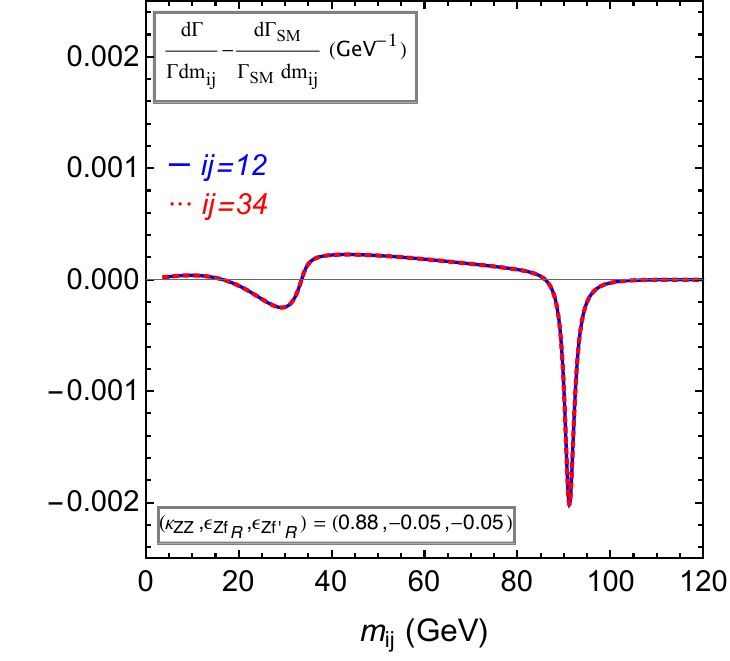} \\
        (a) & (b) \\
    \end{tabular}
   
  \includegraphics[width=0.54\textwidth]{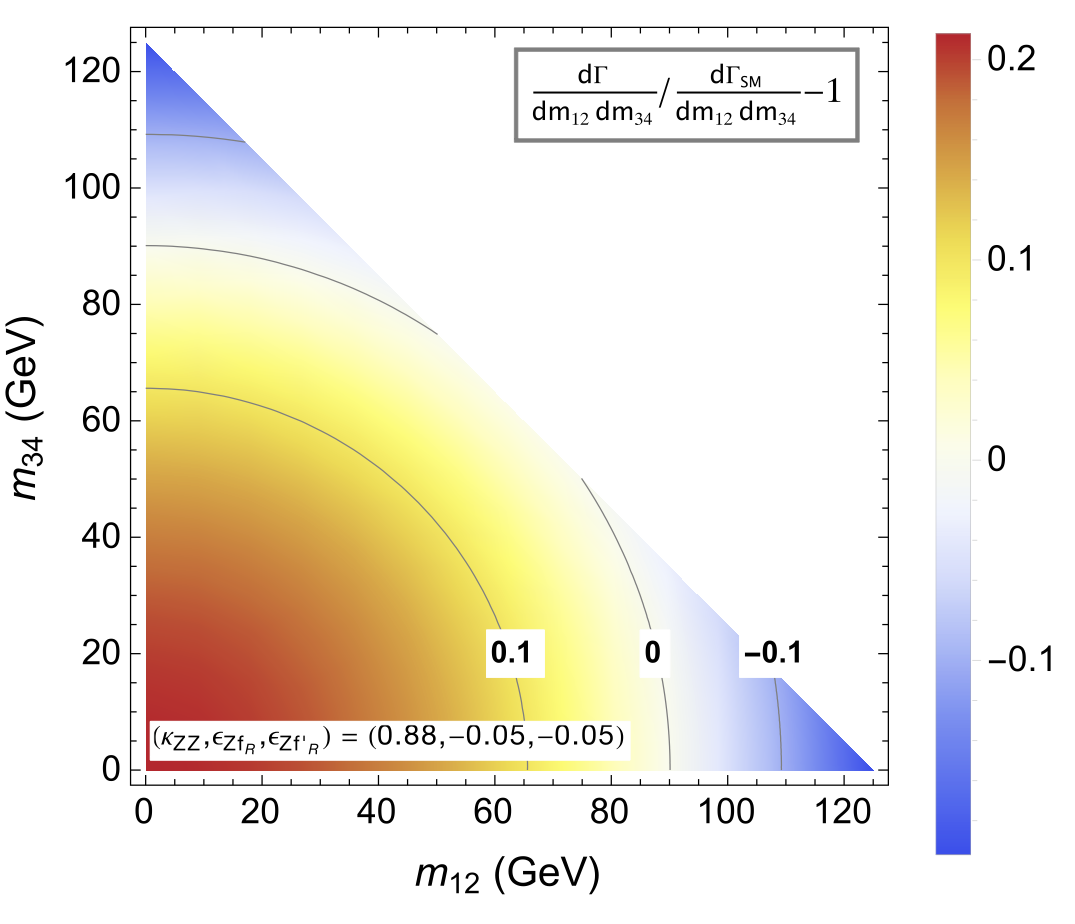} \\ (c)
  
  \caption{\label{fig:tot2} The same plots as in Fig.~\ref{fig:tot} for the case $\epsilon_{Zf_R}=\epsilon_{Zf'_R}$.  The benchmark point in the plots (b) and (c) is $(\kappa_{ZZ},\epsilon_{Zf_R},\epsilon_{Zf'_R})=(0.88,-0.05,-0.05).$}
\end{figure}

\subsection{Measuring contact terms}
\label{sec:ExpContTerms}

In order to probe the contact terms $\epsilon_{Zf}$ and  $\epsilon_{Zf'}$ it is mandatory to exploit the differential 
decay distributions in $q_1^2$ and $q_2^2$.
As an illustration, let us  consider first the case of sizable deviations in one of the $\epsilon_{Zf_R}$ and in $\kappa_{ZZ}$, 
while keeping other couplings SM-like. The ratio of the total Higgs decay rate to $e^+e^- \mu^+\mu^-$ 
with respect to the SM prediction as a function of the couplings is shown in Fig.~\ref{fig:tot} (a).
As can be seen, a measurement of the total rate alone is not capable of resolving the contribution from 
the contact terms.
On the contrary, in Fig.~\ref{fig:tot} (b) we show the deviations from the SM
in the normalized single differential distributions in $m_{12}\equiv \sqrt{q_1^2}$ and $m_{34}\equiv \sqrt{q_2^2}$ in solid-blue line and dashed-red line, respectively. These are obtained after fully integrating Eq.~\eqref{eq-diff-dis} over the corresponding invariant mass. As a benchmark, we set $(\kappa_{ZZ},\epsilon_{Zf_R})=(0.88,-0.10)$,
for which the total decay rate remains as in the SM.   
A good discriminating variable would be the difference between the two distributions. 
This measurement would mainly probe $\epsilon_{Zf_R}$
and provide a complementary information to the one sketched in Fig.~\ref{fig:tot} (a).
Finally, the ratio of the double differential distribution with the SM prediction is 
shown in Fig.~\ref{fig:tot} (c).

Qualitatively, the same discussion holds if both $\epsilon_{Zf_L}$ and $\epsilon_{Zf_R}$
are present, except for a trivial rescaling in the magnitude of the effects. For instance,
if $\epsilon_{Zf_R}=\epsilon_{Zf_L}$ the difference in the differential distributions is
rescaled by the factor $g^f_V/g^f_R$.

Somewhat different signatures are obtained if both $\epsilon_{Zf_R}$ and $\epsilon_{Zf'_R}$
are sizable. As an example, in Fig.~\ref{fig:tot2} we consider the case $\epsilon_{Zf_R}=\epsilon_{Zf'_R}$ 
which corresponds to the relation imposed by flavor universality (See Section~\ref{sec:symmetry}).  
Similarly to case analyzed in Fig.~\ref{fig:tot},  the deviations from the SM predictions are reported. 
As can be seen, the overall size of the effect is much smaller. 
As expected, in this case the single differential distributions in $m_{12}$ and $m_{34}$ are 
the same, and the double differential distribution is symmetric under $m_{12}\leftrightarrow m_{34}$.

\section{Conclusions}

The experimental precision on the Higgs decay distributions, especially those into four light leptons, 
is expected to significantly improve in the next few years. This will allow us to   investigate in depth 
a wide class of possible extensions of the SM. However, to reach this goal, an accurate and sufficiently general 
parameterization of possible NP effects in such distributions is needed. 
  
In this paper we have identified the complete set of pseudo-observables appearing in on-shell Higgs-decay 
distributions in the limit of heavy NP. More precisely,  we only assumed  that contributions to the decay amplitudes generated by effective operators of $D >6$ in a generic EFT approach can be neglected.
The pseudo-observables we have introduced are defined by the momentum expansion of the on-shell Higgs decay amplitudes. 
As such, they are well-defined physical parameters that  can be directly extracted from data, 
providing a natural generalization of the so-called 
``$\kappa$-framework". They indeed consist of four universal ``$\kappa$-like" pseudo-observables ($\kappa_{ZZ}$,
$\kappa_{Z\gamma}$, $\kappa_{\gamma\gamma}$, $\kappa_{WW}$), whose expectation is 1 within the SM,
 and a series of $\epsilon_X$ parameters, whose SM expectation is zero for all practical purposes~(i.e.~it
 is well below the experimental sensitivity even in the HL-LHC era). The ``$\kappa$-like" observables  
 differ from the signal strength measurements currently reported by ATLAS and CMS, being associated to a well-defined 
 (SM-like) kinematical distribution: they describe the (channel-independent) effective couplings of the Higgs 
 boson to the SM gauge fields.  
  The $\epsilon_X$ terms encode possible non-SM effects in the kinematical 
 distributions as well as violations of the accidental SM symmetries. The complete list of the pseudo-observables 
for the Higgs decays to four leptons is reported in Table~\ref{tab:POsumm}: it ranges from a maximum of 20 independent terms,
if no additional symmetry assumption is made, to a minimum of 7 terms under the hypotheses of 
CP-invariance, lepton-flavor universality and custodial symmetry. 

As outlined in Section~\ref{sec:decomposition},  this formalism is well suited to describe all $h\to 4 f$ decay modes: the only difference 
between leptonic, hadronic, and semi-leptonic modes (such as $h\to 2\ell 2q$), is  the list of  $\epsilon_{V f}$  
parameters  ($V=W,Z$) contributing to the given set of decay channels.  In principle, the same formalism (and the same set of pseudo-observables) 
can also describe in general terms NP effects (with non-trivial kinematical distortions) 
in the Higgs production cross-sections controlled by the 
correlation-function in Eq.~\eqref{eq:ThreePointFunction}, namely $\sigma(pp\to hV)$ and the vector-boson fusion process. 
However, in this case more dynamical assumptions are needed due to the possible break-down of the momentum expansion at large energies (see e.g.~Ref.~\cite{Biekoetter:2014jwa}). 
This problem is absent in the Higgs decay amplitudes discussed in this work, where the energy scale is set by $m_h$.

Comparing to existing experimental and phenomenological analyses of $h\to 4 \ell$ decays,
the main difference due to the use of the complete set of pseudo-observables 
is related to the $\epsilon_{Vf}$ terms , which encode the contributions 
generated by $hVf\bar f$ effective contact interactions~\cite{Isidori:2013cla}.
As pointed out in Ref.~\cite{Isidori:2013cga}, 
such terms are particularly interesting in order to discriminate from  data  the hypotheses of linear vs.~non-linear EFT expansion.  This is so because the linear approach predicts relations between electroweak observables and 
$hVf\bar f$ contact terms, leading to strong (and potentially falsifiable) bounds on the latter.
 As we have shown by means of the explicit calculation of the Higgs pseudo-observables in terms of EFT Wilson coefficients,
the linear EFT approach also predicts definite relations among Higgs pseudo-observables, so that not all of them are independent. 
An experimental check of these relations, which involves only Higgs-physics data,  
 would therefore offer an independent tool to 
 possibly discriminate between the linear and the non-linear EFT expansions.  
 
A further interesting aspect of the contact terms (or the $\epsilon_{Vf}$ pseudo-observables) is their 
potential flavor non-universal nature.  Their experimental determination is therefore 
an interesting way to test, from data, the assumption of  flavor-universality 
in the Higgs sector (that is often assumed to hold, up to small breaking terms related to fermion masses).
As we have illustrated with a few examples in the $h\to e^+e^- \mu^+\mu^-$ case, 
the extraction of such terms from data require non-trivial kinematical studies, but significant bounds
could be obtained in the future with high-statistics data. 

\medskip 
Summarizing, the framework of Higgs pseudo-observables provided in this work can capture all the physics accessible in Higgs decays if no new light state is coupled to the $h(125)$ boson; it can be systematically improved with higher-order QCD and QED corrections, recovering the best up-to-date SM predictions in absence of new physics; it can be generalized in a simple way in order to describe any on-shell Higgs decay; it can be efficiently used to test the symmetries of the new-physics sector without specifying the details of the underlying Lagrangian.  We advocate the use of 
such formalism in the era of precise Higgs-boson physics, 
in order to shed light in a systematic and unbiased way on 
the structure and symmetries of possible extensions of the SM.

\subsection*{Acknowledgements}
We thank Michael Duehrssen-Debling, Adam Falkowski, Andre Tinoco Mendes, and Michael Trott for useful discussions.  We also thank Ilaria Brivio for comments on the manuscript. This project was partially funded by the Lyon Institute of Origins, grant ANR-10-LABX-66 (M.G.A.). \vspace{1em} \\ 
\noindent {\bf Note Added.}
While this work was in its final stage, Ref.~\cite{Pomarol:2014dya} appeared, where relations analogous to Eqs.~(\ref{eq:custkVV},\ref{eq:custContL}) have been reported.


\appendix


\section{Matching to the linear EFT}
\label{app:EFTmatch}

In this Appendix we present the expressions of the pseudo-observables defined in Section~\ref{sec:decomposition} in 
terms of the Wilson coefficients of the so-called linear EFT, employing the basis of Ref.~\cite{Pomarol:2013zra}. 
Although most of the details about the EFT can be found in that work, it is worth clarifying a few points
\begin{itemize}
\item Since our flavor symmetry ($\U(1)_f$ for each light fermion) is smaller than the $\U(3)^5$ symmetry of Ref.~\cite{Pomarol:2013zra}, we need to keep fermion indexes in the Wilson coefficients. Moreover we need to keep ${\cal O}_{L}^{\ell}$ and ${\cal O}_{L}^{(3)\ell}$ in the basis.\footnote{~One flavor component of each of these operators is redundant. We choose $[c_{L}^{(3)\ell}]_{ee} = [c_{L}^{\ell}]_{ee} = 0$.}
\item Both these flavor symmetries imply that the 8 dipole operators and ${\cal O}_R^{ud}$ should be suppressed by the Yukawa couplings, and therefore  can be safely neglected.
\item Following Ref.~\cite{Pomarol:2013zra} we define the Wilson coefficients of the 18 operators relevant for us as follows
\ba
{\cal L}_{EFT}^{(D=6)} = \sum_a \frac{c_a}{v^2}{\cal O}_a + \sum_b \frac{\kappa_b}{m_W^2}{\cal O}_b + \sum_{V=B,W} \frac{c_V}{m_W^2}{\cal O}_V ~,
\ea
where $c_a = c_T, c_H, c_R^{u,d,e}, c_L^{q,\ell}, c_L^{(3) q,\ell}, c_{LL}^{3\ell}$,
and $\kappa_b = \kappa_{HB,HW,BB,H\tilde{B},H\tilde{W},B\tilde{B}}$.
The definition of the operators themselves can be found in Ref.~\cite{Pomarol:2013zra}.\footnote{
Under our flavor symmetry assumptions, the coefficient $c_{LL}^{3\ell}$ contains two allowed flavor structures. Instead, we will follow the usual convention of keeping both ${\cal O}_{LL}^{3\ell}$ and ${\cal O}_{LL}^\ell$ but allowing only for one flavor structure, namely $c_{ijkl}=\alpha_{ik}\delta_{ij}\delta_{kl}$.
}
\item Including 24 additional four-fermion operators and ${\cal O}_{6,y_{u,d,e},GG,3W,3G,G\tilde{G},3\tilde{W},3\tilde{G}}$ one recovers
the complete list of  62 flavor-dependent  operator structures, which reduces to 59 independent terms in the one-family case 
 (see Ref.~\cite{Elias-Miro:2013mua} for more details).
\end{itemize}
 
Working at tree-level and at linear order in the NP corrections, we find the following results\footnote{
Notice that Ref.~\cite{Pomarol:2013zra} use $g_f^Z$ for a different quantity, namely $g_f^{Z,SM}\!/2$ in our notation, and that $\delta m_W^2$ is defined in that work with the opposite sign.}
\ba
\kappa_{ZZ} &=& 1 - \frac{c_H}{2} - 2 c_T - \frac{\delta G_F}{2G_F} - \frac{\delta m_Z^2}{m_Z^2} + 2 s_w^2 \left( c_W + c_B \right) +2 \epsilon_{Z\partial Z}~, \label{eq:kZZEFT} \\
\kappa_{WW} &=& 1 - \frac{c_H}{2} - \frac{\delta G_F}{2 G_F}  - \frac{\delta m_W^2}{m_W^2}  + 2\epsilon_{W\partial W}~,\label{eq:kWWEFT} \\
\epsilon_{Z f_L^i} &=&  \frac{2m_Z}{v} \left( T_L^3 [c_L^{(3) f}]_{ii} - \frac{1}{2} [c_L^f]_{ii} \right) + g_{f_L}^Z \epsilon_{Z\partial Z} + eQ_{f_L}   \epsilon_{Z\partial \gamma}  ~~~~~ (f=e,\nu,d) ~,\\
\epsilon_{Z u_L^{ij}} &=&  \frac{2m_Z}{v} \left( T_L^3 (V_{ik} [c_L^{(3)q}]_{kk} V^\dagger_{kj}) - \frac{1}{2} (V_{ik}[c_L^{q}]_{kk} V^\dagger_{kj})  \right)+ g_{u_L}^Z \epsilon_{Z\partial Z} + e Q_{u_L}   \epsilon_{Z\partial \gamma}  ~,\\
\epsilon_{Z f_R^i} &=&  -\frac{m_Z}{v} [c_R^f]_{ii}+ g_{f_R}^Z \epsilon_{Z\partial Z} + eQ_{f_R}   \epsilon_{Z\partial \gamma} ~~~~~~~~~~~~~~~(f=e,u,d) ~,\\
\epsilon_{W e^i} &=& \frac{\sqrt{2} m_W}{v} [c_L^{(3)\ell}]_{ii} + g_W^{\ell} \epsilon_{W\partial W}~,\\
\epsilon_{Wu^i_L d^j_L} &=& \frac{\sqrt{2} m_W}{v} (V_{ij} [c_L^{(3)q}]_{jj} ) + g_W^{ud} \epsilon_{W\partial W}~,\\
\epsilon_{Wu^i_R d^j_R}  &=& 0~, \\
\epsilon_{ZZ} &=&  4 c_{ZZ}~,	\qquad\qquad\qquad		\epsilon_{ZZ}^{CP} =  4 c_{Z\tilde{Z}}~,\\
\epsilon_{Z\gamma} &=& -4 t_w\kappa_{Z\gamma}^{\rm PR} ~,	\qquad\qquad		\epsilon_{Z\gamma}^{CP} =  -4 t_w\kappa_{Z\tilde{\gamma}}~,\\
\epsilon_{\gamma\gamma} &=& -8 s_w^2 \kappa_{BB} ~,	\qquad\qquad		\epsilon_{\gamma\gamma}^{CP} =  -8 s_w^2 \kappa_{B\tilde{B}}~,\\
\epsilon_{WW} &=& 2\kappa_{HW}~,	\qquad\qquad		\epsilon_{WW}^{CP} =  2\kappa_{H\tilde{W}}~.
\ea
where $V$ is the CKM matrix\footnote{We define our flavor symmetry in the basis where the down-quark and charged lepton Yukawa matrices are diagonal, whereas the up-quark Yukawa matrix has the form $Y_U=V^\dagger Y_U^{\rm diag}$. We neglect the breaking of the symmetry induced by the Yukawa matrices
but for its effect on fermion masses. 
See Ref.~\cite{Cirigliano:2009wk} (Section~3) for a more detailed discussion.}
(also notice that $[c_L^{(3) \ell}]_{ee}=[c_L^\ell]_{ee}=0$ in our basis)
and where the $\epsilon_{V\partial V}$ coefficients, introduced in Eq.~\eqref{eq:hVVEFTFormFactors2}, are given by
\be
	\epsilon_{Z\partial Z} = \hat c_{Z} ~, \qquad
	\epsilon_{Z\partial \gamma} = t_w (\hat c_{W}-\hat c_B)~, \qquad
	\epsilon_{W\partial W} = \hat c_{W}~.
\ee
Also, $\kappa_{Z \gamma}^{\rm PR}$, $c_{ZZ}$ and $\hat c_{Z/W/B}$ are the combination of Wilson coefficients defined in Ref.~\cite{Pomarol:2013zra}, namely
\ba
&&
\hat c_{Z}=\hat c_{W}+ \hat c_B t^2_w \, \qquad
\hat c_{W}= c_W+\kappa_{HW}\ , \qquad
\hat c_{B}=c_B+\kappa_{HB}\, , 
\\
&&
c_{ZZ}=\frac{1}{2} (\kappa_{HW} +\kappa_{HB} t^2_w)-2 \frac{s_w^4}{c_w^2} \kappa_{BB}\, .\label{cwwczz}~,\\
&&
\kappa_{Z\gamma}^{\rm PR} =-\frac{1}{4} \left( \kappa_{HW} - \kappa_{HB} \right) - 2s_w^2 \kappa_{BB} ~, 
\ea
and likewise for the CP-odd terms. The contributions to the pseudo-observables proportional to the $\epsilon_{Z\partial Z}$, $\epsilon_{W\partial W}$ and $\epsilon_{Z \partial \gamma}$ coefficients are due to the redefinition of Eqs.~(\ref{eq:kappaRedef},\ref{eq:contIntCustDecomp}), which are necessary in order to match with our pseudo-observables.

The $\kappa_{ZZ}$ and $\kappa_{WW}$ parameters are the only ones already present in the SM at tree level. For this reason they can receive contributions from the EFT either directly from $D=6$ operators, such as the terms proportional to $c_H$ and $c_T$ in Eqs.~(\ref{eq:kZZEFT},\ref{eq:kWWEFT}), or via a rescaling in the kinetic term (like the contribution proportional to $c_W + c_B = \hat{S}$), or finally through a variation of the SM input parameters.
In particular, the terms $\delta m_{Z,W}^2$ and $\delta G_F$ contain the NP contributions that contaminate the determination of the SM parameters $(g, g^\prime, v)$ from the measurement of some input observables, and indirectly affect the pseudo-observables $\kappa_{ZZ,WW}$ through the $m_{Z,W}^2/v_F$ term of Eqs.~\eqref{eq:h4l1} and ~\eqref{eq:h4lCharged}. A common set of input observables used to fix the SM parameters includes the $Z$-boson mass, the low-energy fine-structure constant, $\alpha_{\rm em}(0)$, and $G_F$ extracted from the muon lifetime. The experimental value of the $Z$-boson mass and $\alpha_{\rm em}$ are modified by the following $D=6$ effective operators
\be
	 \frac{\delta m_Z^2}{m_Z^2} = - c_T + 2 s_w^2 \left( c_W + c_B \right)~, \qquad
	\frac{\delta \alpha_{\rm em}}{\alpha_{\rm em}} = - 2 s_w^2 \left( c_W + c_B \right)~,
\ee
whereas the $G_F$ determination from the muon lifetime is changed by~\cite{Cirigliano:2009wk}
\ba
\frac{\delta G_F}{G_F} = - 2 [c_{LL}^{(3) \ell}]_{ee\mu\mu} \,+\, [c_L^{(3)\ell}]_{\mu\mu}~,
\ea
where we have used that $[c_{L}^{(3)\ell}]_{ee} = 0$ in our basis. 
For this choice of input observables, the variation of the $W$ mass is given by
\ba
\label{eq:mWderived}
\frac{\delta m_W^2}{m_W^2} = \frac{1}{c_{2w}} \left[ s_w^2 \left(2c_W + 2c_B + \frac{\delta G_F}{G_F}\right) - c_w^2 c_T \right]~. \qquad
\ea
If, instead of $G_F$ from the muon lifetime (or of $\alpha_{em}(0)$), we use the experimental measurement of $m_W$ as an input observable, then $\delta m_W^2$ vanishes.

\subsection[Checking custodial symmetry relations]{Checking custodial symmetry relations [Eqs.~(\ref{eq:custVV}-\ref{eq:custContR})]}
\label{app:EFTmatchCust}

From the results presented above it is straightforward to check that the following three relations are satisfied
\ba
&&c_W^2 \epsilon_{ZZ} + 2 c_W s_W \epsilon_{Z\gamma} + s_W^2 \epsilon_{\gamma \gamma} - \epsilon_{WW} = 0~,\\
&& c_W^2 \epsilon^{CP}_{ZZ} + 2 c_W s_W \epsilon^{CP}_{Z\gamma} + s_W^2 \epsilon^{CP}_{\gamma \gamma} - \epsilon^{CP}_{WW} = 0~,\\
&&\epsilon_{W e^i} - \frac{c_W}{\sqrt{2}} (\epsilon_{Z \nu_L^i} - \epsilon_{Z e_L}) =0 ~.
\ea
As explained in the text, this is the consequence of an accidental custodial symmetry in the corresponding $D=6$ operators in the linear EFT case. Concerning the relation~\eqref{eq:custkVV} we find
\be\begin{split}
\kappa_{WW} - \kappa_{ZZ} +& \frac{2}{g} \left( \sqrt{2} \epsilon_{W e_L^i} + 2 c_W \epsilon_{Z e_L^i} \right) = 
- 2 [c_L^{\ell}]_{ii} + c_T - \frac{\delta m_W^2}{m_W^2}\\
&= - 2 [c_L^{\ell}]_{ii} + c_T \frac{2-3 s_w^2}{c_2w}  - \frac{s_w^2}{c_{2w}} \frac{\delta G_F}{G_F} - \frac{2 s_w^2}{c_{2w}} (c_W + c_B)\\
&\stackrel{g'\to 0}{\to}~ - 2 [c_L^{\ell}]_{ii} + 2 c_T~.
\label{eq:CustRel_kappa_EFT}
\end{split}\ee
Both operators ${\cal O}_T$ and ${\cal O}_{L, ii}^{\ell}$ break custodial symmetry~\cite{Elias-Miro:2013mua}. Thus the r.h.s of Eq.~\eqref{eq:CustRel_kappa_EFT} vanishes if the new physics is custodially invariant, confirming Eq.~\eqref{eq:custkVV}. Here we have used Eq.~\eqref{eq:mWderived} for $\delta m_W$. Notice that by using instead the experimental value of $m_W$ as an input, the limit $g' \rightarrow 0$ is not necessary in order for Eq.~\eqref{eq:custContL} to be satisfied since in this case $\delta m_W^2 = 0$.
Finally, for the relation~\eqref{eq:custContR} we find
\be
\epsilon_{Z e_R^i} - \epsilon_{Z \nu_L^i} - \epsilon_{Z e_L^i} = \frac{2m_Z}{v} [c_L^{\ell}]_{ii} - \frac{m_Z}{v} [c_R^{e}]_{ii}~. \label{eq:CustRel_eR_EFT}
\ee
Once again $[c_L^{\ell}]_{ii}$ vanishes if custodial symmetry is imposed, whereas the behavior of the second coefficient, $[c_R^{e}]_{ii}$, depends on the embedding of the right-handed electron. In the case $(A)$, in which $e_R \sim ({\bf 1}, {\bf 2})_{-\frac{1}{2}}$, Eq.~\eqref{eq:custContR} is not expected to be satisfied and in fact it is not since the operator ${\cal O}_R^e$ doesn't break the symmetry, transforming as a singlet. In the case of the embedding $(B)$, where $e_R \sim  ({\bf 1}, {\bf 1})_{-1}$, ${\cal O}_R^e$ transforms as a triplet of the custodial symmetry \cite{Elias-Miro:2013mua} and therefore $c_R^e \neq 0$ is an explicit breaking, so that in the custodially symmetric limit one indeed recovers Eq.~\eqref{eq:custContR}.


\section{Custodial symmetry}
\label{app:Custodial}

In this Appendix we provide an extended discussion on the custodial symmetry relations among the pseudo-observables in Higgs decays.
We are assuming that the new physics sector enjoys a global symmetry $G = \SU(2)_L \times \SU(2)_R \times \U(1)_X$, spontaneously broken to the custodial subgroup $H = \SU(2)_{L+R} \times \U(1)_X$ by the vev of some field $U \sim (2,2)_0$, $\langle U \rangle = {\bf 1}_2$.
Since the hypercharge gauge boson and the SM fermions are not in complete representations of G, their couplings with the BSM sector (i.e. $g^\prime$ and the Yukawa couplings) break the symmetry explicitly.
An efficient way to keep track of the effects of these breaking terms is to promote SM multiplets to complete representations of G by introducing spurion (unphysical) fields which are then set to zero in physical processes.
In the gauge sector, we introduce spurion gauge bosons, so that the whole group $G$ is gauged. We thus introduce the gauge fields $L_\mu^a, R_\mu^a, X_\mu$ and couplings $g, \tilde g, g_X$, respectively for the factors $\SU(2)_L,~ \SU(2)_R,~ \U(1)_X$ (note that in general the two $\SU(2)$ factors can have different coupling). The SM gauging is obtained by setting $L_\mu^a = W_\mu^a$, $R_\mu^a = \delta^{a3} c_X B_\mu$ and $X_\mu = s_X B_\mu$, where
\be
	c_X = \frac{g_X}{\sqrt{\tilde g^2 + g_X^2}} = \frac{g^\prime}{\tilde g} < 1~, \qquad
	s_X = \frac{\tilde g}{\sqrt{\tilde g^2 + g_X^2}} = \frac{g^\prime}{g_X} < 1~.
	\label{eq:GaugeCouplRelations}
\ee
Since the fields $R_\mu^3$ and $X_\mu$ enter in interactions always with the combinations $\tilde g R_\mu^3 = g^\prime B_\mu$ and $g_X X_\mu = g^\prime B_\mu$, we are free to choose any values of $\tilde g, g_X$, provided Eq.~\eqref{eq:GaugeCouplRelations} is satisfied. In particular it is possible to choose $\tilde g = g$,  such that 
 $g_X = g^\prime g / \sqrt{g^2 - g^{\prime 2}}$ as in Ref.~\cite{Contino:2013kra}.
 The hypercharge is given by $Y = T_R^3 + X$ and the electromagnetic charge is then given by $Q = T_L^3 + Y$ (where $T^3_{L,R} = \sigma^3/2$).

\subsection{Fermion embedding}

We also assume that all SM fields couple only to one BSM operator each, so that we can assign them univocal $T_{L,R}$ and $T_{L,R}^3$ quantum numbers depending on the operator they couple to \cite{Agashe:2006at}.
This fixes the representation of G in which we embed the SM fermions.
We focus on leptons and consider only two simple embeddings, see App.~C of Ref.~\cite{Elias-Miro:2013mua}. The first one is
\be
(A)\qquad
\begin{array}{c|c|c|c}
					& \SU(2)_L & \SU(2)_R & \U(1)_X  \\ \hline
L_L = (\nu_L, e_L)^t	&	{\bf 2} &	{\bf 1}   &	- 1/2 \\
L_R = (0, e_R)^t	&	{\bf 1} &	{\bf 2}   &	- 1/2 \\
\end{array}~,
\label{eq:FermEmbeddingA}
\ee
The second embedding we consider is\footnote{~The embedding of $L_L$ in the bidoublet can be explicitly realized in a basis of $2\times 2$ matrices as $E_L = \sigma^\alpha E_L^\alpha$. In particular we have $E_L = \sigma^+ \nu_L + \sigma^{0-} e_L$, where $\sigma^{\pm} = (\sigma^1 \pm \sigma^2)/2$ and $\sigma^{0\pm} = ({\bf 1}_2 \pm \sigma^3)/2$, such that $T^3_{L}(\nu_L) = -T^3_{R}(\nu_L) = T^3_{L}(e_L) = T^3_{R}(e_L) = 1/2$.}
\be
(B)\qquad
\begin{array}{c|c|c|c}
					& \SU(2)_L & \SU(2)_R & \U(1)_X  \\ \hline
E_L 	&	{\bf 2} &	{\bf 2}   &	- 1 \\
e_R	&	{\bf 1} &	{\bf 1}   &	- 1 \\
\end{array}~,
\label{eq:FermEmbeddingB}
\ee
Let us also introduce the fermionic currents which couple to the spurion custodial gauge bosons:
\be
	\mathcal{L} \supset g J_L^{\mu a} L_\mu^a + \tilde g J_R^{\mu a} R_\mu^a + g_X (J_{LX}^\mu + J_{RX}^\mu) X_\mu~.
\ee
where, considering for simplicity the case of one generation only, we have
\be \begin{split}
	J_{LX}^{\mu} &= X_{\ell_L} \bar{L}_L \gamma^\mu L_L~,\\
	J_{RX}^{\mu} &= X_{e_R} \bar{e_R} \gamma^\mu e_R~,\\
	(A)\qquad
	J_L^{\mu a} &= \bar{L}_L \frac{\sigma^a}{2} \gamma^\mu L_L~, \quad
	J_R^{\mu a} = \bar{L}_R \frac{\sigma^a}{2} \gamma^\mu L_R~, \\
	(B)\qquad
	J_L^{\mu a} &= \text{Tr}\left[\bar{E}_L \frac{\sigma^a}{2} \gamma^\mu E_L\right]~, \quad
	J_R^{\mu a} = 0.
	\label{eq:Currents}
\end{split}\ee

In the rest of this section we obtain the most generic form of 1PI Green functions of a Higgs coupling with two gauge bosons and with one gauge boson and one fermonic current.
Of course, this classification is not physical since, by using the equations of motion, it is possible to exchange some $h VV$ interactions for some contact terms $h V J$, and viceversa, as we show below.

\subsection{$hVV$ interactions}

We first 
review here the derivation of the custodial symmetry relation for the interactions of a Higgs with two EW gauge fields, Eq.~\eqref{eq:CustRelFFVV2}, 
following Ref.~\cite{Contino:2013kra}. Given the symmetry breaking pattern $G \rightarrow H$,
the $hVV$ interactions are fully characterized by four form factors:
\be\begin{split}
	\langle h | L^{a\mu}(q_1) L^{a\nu}(q_2) \rangle = i \Gamma_{LL}^{\mu\nu}(q_1, q_2)~, \quad &
	\langle h | L^{a\mu}(q_1) R^{a\nu}(q_2) \rangle = i \Gamma_{LR}^{\mu\nu}(q_1, q_2)~, \\
	\langle h | R^{a\mu}(q_1) R^{a\nu}(q_2) \rangle = i \Gamma_{RR}^{\mu\nu}(q_1, q_2)~, \quad &
	\langle h | X^\mu(q_1) X^\nu(q_2) \rangle = i \Gamma_{XX}^{\mu\nu}(q_1, q_2)~,
\end{split}\ee
or, in other words, by the effective Lagrangian (in momentum space)
\be\begin{split}
	\cL_{eff}^{hVV} =& h \left( \frac{1}{2} L_{\mu}^a(q_1) L_{\nu}^a(q_2) \Gamma^{\mu\nu}_{LL} (q_1, q_2) + L_{\mu}^a(q_1) R_{\nu}^a(q_2) \Gamma^{\mu\nu}_{LR} (q_1, q_2) + \right. \\
		& \left. + \frac{1}{2} R_{\mu}^a(q_1) R_{\nu}^a(q_2) \Gamma^{\mu\nu}_{RR} (q_1, q_2) + \frac{1}{2} X_{\mu}(q_1) X_{\nu}(q_2) \Gamma^{\mu\nu}_{XX} (q_1, q_2) \right)~.
\end{split}\ee
By switching off the unphysical fields we get three independent form factors for the Higgs interactions with the SM gauge bosons:
\be\begin{split}
	& \langle h | W^{a\mu}(q_1) W^{a\nu}(q_2) \rangle = i \Gamma_{LL}^{\mu\nu}(q_1, q_2)~, \qquad
	\langle h | W^{3\mu}(q_1) B^{\nu}(q_2) \rangle = i c_X \Gamma_{LR}^{\mu\nu}(q_1, q_2)~, \\
	& \langle h | B^\mu(q_1) B^\nu(q_2) \rangle = i c_X^2 \Gamma_{RR}^{\mu\nu}(q_1, q_2) + i s_X^2 \Gamma_{XX}^{\mu\nu}(q_1, q_2) \equiv i \Gamma_{BB}^{\mu\nu}(q_1, q_2)~.
\end{split}
\label{eq:FormFactorsGauge}\ee
In particular, the distinction between the $XX$ and the $RR$ form factors is not physical.
Let us note that, while imposing only $\U(1)_{\rm em}$ invariance the $\langle h | W_\mu^3 W_\nu^3 \rangle$ and $\langle h | W_\mu^+ W_\nu^- \rangle $ form factors are independent, custodial symmetry relates both of them to the $\langle h | W_\mu^a W_\nu^a \rangle$ one.
In fact, the generic $\U(1)_{\rm em}$-invariant 1PI Green functions describing the couplings of a Higgs with two SM EW gauge bosons are:
\be\begin{split}
	\langle h | W^{+\mu}(q_1) W^{-\nu}(q_2) \rangle = i \Gamma_{WW}^{\mu\nu}(q_1, q_2)~, \quad &
	\langle h | Z^\mu(q_1) Z^\nu(q_2) \rangle = i \Gamma_{ZZ}^{\mu\nu}(q_1, q_2)~, \\
	\langle h | Z^\mu(q_1) A^\nu(q_2) \rangle = i \Gamma_{Z\gamma}^{\mu\nu}(q_1, q_2)~, \quad &
	\langle h | A^\mu(q_1) A^\nu(q_2) \rangle = i \Gamma_{\gamma\gamma}^{\mu\nu}(q_1, q_2)~.
	\label{eq:PhyshVVFormFact}
\end{split}\ee
Since, in a custodially-invariant theory these form factors arise from the three in Eq.~(\ref{eq:FormFactorsGauge}), they are not independent \cite{Contino:2013kra}: 
\be
	\Gamma_{WW}^{\mu\nu}(q_1, q_2) = c_w^2 \Gamma_{ZZ}^{\mu\nu}(q_1, q_2) + c_w s_w (\Gamma_{Z\gamma}^{\mu\nu}(q_1, q_2) + \Gamma_{Z\gamma}^{\nu\mu}(q_2, q_1)) + s_w^2 \Gamma_{\gamma\gamma}^{\mu\nu}(q_1, q_2)~.
	\label{eq:CustRelFFVV2}
\ee
By expanding the form factors in powers of momenta over the cutoff of the EFT up to $D=6$ terms one has
\be\begin{split}
	\Gamma_{VV}^{\mu\nu}(q_1, q_2) =& \frac{2 m_V^2}{v_F} \left( \kappa^0_{VV} g^{\mu\nu} + \frac{\epsilon_{VV}}{m_V^2}  P_T^{\mu\nu}(q_1,q_2) + \frac{\epsilon_{VV}^{CP}}{m_V^2} \epsilon^{\mu\nu\rho\sigma} q_{2\rho} q_{1\sigma} \right)~+ \\
		&+ \frac{2}{v_F} \epsilon_{V\partial V} (P_D^{\mu\nu}(q_1) + P_D^{\mu\nu}(q_2))~, \quad (V = W,Z)\\
	\Gamma_{Z\gamma}^{\mu\nu}(q_1, q_2) =& \frac{2}{v_F} \epsilon_{Z\gamma}  P_T^{\mu\nu}(q_1,q_2) + \frac{2 \epsilon_{Z\gamma}^{CP}}{v_F} \epsilon^{\mu\nu\rho\sigma} q_{2\rho} q_{1\sigma} + \frac{2}{v_F} \epsilon_{Z\partial \gamma} P_D^{\mu\nu}(q_2)~,\\
	\Gamma_{\gamma\gamma}^{\mu\nu}(q_1, q_2) =& \frac{2}{v_F} \epsilon_{\gamma\gamma}  P_T^{\mu\nu}(q_1,q_2) + \frac{2 \epsilon_{\gamma\gamma}^{CP}}{v_F} \epsilon^{\mu\nu\rho\sigma} q_{2\rho} q_{1\sigma}~,
	\label{eq:hVVEFTFormFactors2}
\end{split}\ee
where $P_T^{\mu\nu}(q_1,q_2) = q_1 \cdot q_2 g^{\mu\nu} - q_2^\mu q_1^\nu$ and $P_D^{\mu\nu}(q) = g^{\mu\nu} q^2 - q^\mu q^\nu$. From this expansion, by equating terms in Eq.~\eqref{eq:CustRelFFVV2} with the same momentum dependence, one gets the relations:
\be\begin{split}
	\kappa^0_{WW} &= \kappa^0_{ZZ}~,\\
	\epsilon_{W\partial W} &= c_w^2 \epsilon_{Z\partial Z} + c_w s_w \epsilon_{Z\partial \gamma}~,\\
	\epsilon_{WW} &= c_w^2 \epsilon_{ZZ} + 2 c_w s_w \epsilon_{Z\gamma} + s_w^2 \epsilon_{\gamma \gamma}~,\\
	\epsilon^{CP}_{WW} &= c_w^2 \epsilon^{CP}_{ZZ} + 2 c_w s_w \epsilon^{CP}_{Z\gamma} + s_w^2 \epsilon^{CP}_{\gamma \gamma}~.
	\label{eq:CustRelhVVEFT2}
\end{split}\ee
In order to make a connection between these unphysical coefficients and our pseudo-observables, it is necessary to calculate the amplitude for a physical process involving on-shell particles and match  it  with Eqs.~(\ref{eq:h4l1},\ref{eq:h4lNeutrCurrEFT1}-\ref{eq:h4l2}).
By doing so one recognizes that the $\epsilon_{WW,ZZ,Z\gamma,\gamma\gamma}$ coefficients and their CP-odd counterparts are identical to the analogous pseudo-observables, while some combinations of the coefficients  $\epsilon_{W \partial W,Z \partial Z, Z\partial \gamma}$ and $\kappa_{WW,ZZ}^0$ describe
 contact interactions of the type $h V J$.
This redundancy is easily understood by computing the amplitude for the physical process $h \rightarrow J^{\mu}_{f} J^{\nu}_{f^\prime}$ arising from these couplings:
\be\begin{split}
	&\cA(h \rightarrow J^{+\mu}_{\ell_L}(q_1) J^{- \nu}_{\ell_L}(q_2)) = \frac{2i }{v_F} \left( (q_1^2 + q_2^2) \epsilon_{W\partial W} g_W^{f} g_W^{f^\prime}  + m_W^2 \kappa_{WW}^0 g_W^{f} g_W^{f^\prime} \right)\frac{ g_{\mu\nu}}{P_W(q_1^2)P_W(q_2^2)} J^{+\mu}_{\ell_L} J^{- \nu}_{\ell_L}~,\\
	&\cA(h \rightarrow J^{\mu}_{f}(q_1) J^{\nu}_{f^\prime}(q_2)) =  \frac{2i }{v_F} \left[ \left( (q_1^2 + q_2^2) \epsilon_{Z\partial Z} g_Z^{f} g_Z^{f^\prime}  + m_Z^2 \kappa_{ZZ}^0 g_Z^{f} g_Z^{f^\prime} \right)\frac{ 1}{P_Z(q_1^2)P_Z(q_2^2)} + \right.~\\
	&\qquad\qquad\qquad\qquad\qquad\qquad \left. + q_1^2 \epsilon_{Z\partial \gamma} \frac{e Q_f g_Z^{f^\prime}}{q_1^2 P_Z(q_2^2)} + q_2^2 \epsilon_{Z\partial \gamma} \frac{e Q_{f^\prime} g_Z^{f}}{q_2^2 P_Z(q_1^2)} \right] g_{\mu\nu} J^{\mu}_{f} J^{\nu}_{f^\prime} ~.
\end{split}\ee
For $\kappa^0_{WW} = - 2 \epsilon_{W\partial W} $ and $\kappa^0_{ZZ} = - 2 \epsilon_{Z\partial Z} $ the amplitude has exactly the same structure as the contact interactions in Eq.~\eqref{eq:h4lNeutrCurrEFT1}.\footnote{~In terms of  EFT operators, this redundancy is a consequence of the fact that,  by using the equations of motion, one can rewrite the $h V^\mu D^\nu V_{\mu\nu}$ operators, responsible for the $\epsilon_{V\partial V}$ terms, as a combination of $h V_\mu \bar{f}\gamma^\mu f^\prime$ contact interactions and $m_V^2 h V^\mu V_\mu$ terms.} In order to match with our parametrization, we thus redefine $\kappa^0_{WW,ZZ}$ as follows:
\be
	\kappa^0_{WW} = \kappa_{WW} - 2 \epsilon_{W\partial W}~, \qquad
	\kappa^0_{ZZ} = \kappa_{ZZ} - 2 \epsilon_{Z\partial Z}~.
	\label{eq:kappaRedef}
\ee

Therefore, the contact interactions receive two separate contributions: the \emph{direct} ones from $D\leq6$ operators contributing to $\langle h | J_f^\mu V^\nu \rangle_{\rm 1 PI}$, $\epsilon_{V f}^D$, and the \emph{indirect} ones due to the matching described above:
\be \begin{split}
	\epsilon_{W e_L^i} &= \epsilon_{W e_L^i}^D + \frac{g}{\sqrt{2}} \epsilon_{W\partial W} ~, \\
	\epsilon_{Z e_L^i} &= \epsilon_{Z e_L^i}^D + \frac{g}{c_w} \left(-\frac{1}{2} + s_w^2 \right) \epsilon_{Z\partial Z} - g s_w \epsilon_{Z\partial \gamma}~, \\
	\epsilon_{Z \nu_L^i} &= \epsilon_{Z \nu_L^i}^D + \frac{g}{2 c_w} \epsilon_{Z\partial Z} ~,\\
	\epsilon_{Z e_R^i} &= \epsilon_{Z e_R^i}^D + \frac{g}{c_w} s_w^2 \epsilon_{Z\partial Z} - g s_w \epsilon_{Z\partial \gamma}~.
	\label{eq:contIntCustDecomp}
 \end{split}\ee
The division between direct and indirect contributions is not physical; only their sum is a physical and observable quantity. The indirect contributions above satisfy two 
independent relations, one due to the fact that we have three coefficients $\epsilon_{V\partial V}$ describing four contact terms, and a second one due to  the custodial symmetry relation of Eq.~\eqref{eq:CustRelhVVEFT2}.
It is then convenient to parametrize the observable $\kappa$ coefficients as
\be
	\kappa_{WW} = 1 + \delta \kappa + \delta \kappa_{WZ}~, \qquad
	\kappa_{ZZ} = 1 + \delta \kappa~,
	\label{eq:kappaPhysParam}
\ee
so that after the redefinition of Eq.~\eqref{eq:kappaRedef} we can rewrite the custodial symmetry relation $\kappa_{WW}^0 = \kappa_{ZZ}^0$ as $\delta \kappa_{WZ} = 2 (\epsilon_{W\partial W} - \epsilon_{Z\partial Z})$. 

\subsection{$hVJ$ interactions}

Let us now turn to the direct contribution to contact interactions. Such terms arise from 1PI Green functions of the type $\langle h | J_f^\mu V^\nu \rangle$. Let us study these interactions for the two embeddings introduced above.

\subsubsection*{Embedding $A$}

We start by considering the embedding $A$ of Eq.~\eqref{eq:FermEmbeddingA}. We define the possible 1PI Green functions in a custodially invariant theory in terms of form factors as:
\be\begin{split}
	\langle h | J^{a\mu}_{L }(q_1) L^{a\nu}(q_2) \rangle = i F_{LL}^{\mu\nu}(q_1, q_2)~, \qquad &
	\langle h | J^{a\mu}_{L}(q_1) R^{a\nu}(q_2) \rangle = i F_{LR}^{\mu\nu}(q_1, q_2)~, \\
	\langle h | J^{a\mu}_{R}(q_1) R^{a\nu}(q_2) \rangle = i F_{RR}^{\mu\nu}(q_1, q_2)~, \qquad &
	\langle h | J^{a\mu}_{R}(q_1) L^{a\nu}(q_2) \rangle = i F_{RL}^{\mu\nu}(q_1, q_2)~, \\
	\langle h | J_{LX}^\mu(q_1) X^\nu(q_2) \rangle = i F_{LX}^{\mu\nu}(q_1, q_2)~, \qquad &
	\langle h | J_{RX}^\mu(q_1) X^\nu(q_2) \rangle = i F_{RX}^{\mu\nu}(q_1, q_2)~.
	\label{eq:hvJCustFF}
\end{split}\ee
By switching off the unphysical fields we get the following contact interactions of the Higgs with one fermion current and a SM EW gauge boson:
\be\begin{split}
	\langle h | J^{a\mu}_{\ell_L} W^{a\nu} \rangle &= i F_{LL}^{\mu\nu}~, \\
	\langle h | J^{a\mu}_{e_L } B^\nu \rangle &= i \left( - \frac{1}{2} c_X F_{LR}^{\mu\nu} + X_{\ell_L} s_X F_{LX}^{\mu\nu} \right) \equiv i F_{e_L B}^{\mu\nu} ~, \\
	\langle h | J^{a\mu}_{\nu_L} B^\nu \rangle &= i \left( \frac{1}{2} c_X F_{LR}^{\mu\nu} + X_{\ell_L} s_X F_{LX}^{\mu\nu} \right) \equiv i F_{\nu_L B}^{\mu\nu} ~, \\
	\langle h | J_{e_R}^\mu W^{3\nu} \rangle &= - \frac{1}{2} F_{RL}^{\mu\nu} \equiv i F_{e_R W}^{\mu\nu} ~, \\
	\langle h | J_{e_R}^\mu B^\nu \rangle &= i \left( - \frac{1}{2} c_X F_{RR}^{\mu\nu} + X_{e_R} s_X F_{RX}^{\mu\nu} \right) \equiv i F_{e_R B}^{\mu\nu} ~,
\end{split}
\label{eq:hvJCustFFSM}\ee
where
\be
	J^a_{\ell_L \mu} = \bar{L}_L \frac{\sigma^a}{2} \gamma_\mu L_L, \qquad J_{e_{L,R} \mu} = \bar{e}_{L,R} \gamma_\mu e_{L,R} \quad\text{and}\quad J_{\nu_{L} \mu} = \bar{\nu}_{L} \gamma_\mu \nu_{L}.
	\label{eq:physCurrents}
\ee
In terms of the mass eigenstates there are seven possible contact terms:
\be\begin{split}
	\langle h | J^{+\mu}_{\ell} W^{+\nu} \rangle = i F_{W \ell}^{\mu\nu} = \frac{2i }{v_F} \epsilon_{W \ell}^D g^{\mu\nu}~, \;\; & \\
	\langle h | J_{e_L}^\mu Z^\nu \rangle = i F_{Z e_L}^{\mu\nu} = \frac{2i }{v_F} \epsilon_{Z e_L}^D g^{\mu\nu}~, \quad & \quad
	\langle h | J_{e_L}^\mu A^\nu \rangle = i F_{\gamma e_L}^{\mu\nu} = 0~, \\
	\langle h | J_{\nu_L}^\mu Z^\nu \rangle = i F_{Z \nu_L}^{\mu\nu}=  \frac{2i }{v_F} \epsilon_{Z \nu_L}^D g^{\mu\nu}~,\quad & \quad
	\langle h | J_{\nu_L}^\mu A^\nu \rangle = i F_{\gamma \nu_L}^{\mu\nu} = 0~, \\
	\langle h | J_{e_R}^\mu Z^\nu \rangle = i F_{Z e_R}^{\mu\nu} =  \frac{2i }{v_F} \epsilon_{Z e_R}^D g^{\mu\nu}~,\quad & \quad
	\langle h | J_{e_R}^\mu A^\nu \rangle = i F_{\gamma e_R}^{\mu\nu} = 0~,
	\label{eq:CoeffVJEFT}
\end{split}\ee
where we also provide the EFT expansion up to $D=6$ terms. Note that the vertices of a current with a photon are not present at $D\leq6$ due to the $U(1)_{\rm em}$ invariance, they appear only at $D>6$.
Independently of the EFT expansion, since only five form factors are independent, one has two relations from custodial symmetry:
\be\begin{split}
	F_{W \ell}^{\mu\nu}(q_1,q_2) &= - \sqrt{2} \left( s_w F_{\gamma e_L}^{\mu\nu}(q_1,q_2) +  c_w F_{Z e_L}^{\mu\nu}(q_1,q_2) \right)~, \\
	F_{W \ell}^{\mu\nu}(q_1,q_2) &= \sqrt{2} \left( s_w F_{\gamma \nu_L}^{\mu\nu}(q_1,q_2) +  c_w F_{Z \nu_L}^{\mu\nu}(q_1,q_2) \right)~.
	\label{eq:CustRelFFVJ}
\end{split}\ee
In terms of the EFT coefficients these relations read
\be
	\epsilon_{W \ell}^D = - \sqrt{2} c_w \epsilon_{Z e_L}^D~, \qquad \epsilon_{Z \nu_L}^D = - \epsilon_{Z e_L}^D ~.
	\label{eq:contIntCustEFT2}
\ee
One can notice that in this case custodial symmetry simply implies that the form factors of $\langle h | J_{e_L}^\mu W^{3\nu} \rangle$, $\langle h | J_{\nu_L}^\mu W^{3\nu} \rangle$ and $\langle h | J_{\ell_L}^{\mu+} W^{+\nu} \rangle$ all arise from a single term $\langle h | J_{\ell_L}^{a\mu} W^{a\nu} \rangle$. These conditions are independent of the embedding of the left-handed doublet, in particular they apply also to the embedding $(B)$.
The dependence on the embedding shows up only in the couplings of the right-handed fermions, which in this case remain arbitrary.

\subsubsection*{Embedding $B$}

Let us study now the embedding $B$ of Eq.~\eqref{eq:FermEmbeddingB}. In this case we can write the following two-point functions of a gauge boson and a fermonic current:
\be\begin{split}
	\langle h | J^{a\mu}_{L}(q_1) L^{a\nu}(q_2) \rangle = i F_{LL}^{\mu\nu}(q_1, q_2)~, \qquad &
	\langle h | J^{a\mu}_{L}(q_1) R^{a\nu}(q_2) \rangle = i F_{LR}^{\mu\nu}(q_1, q_2)~, \\
	\langle h | J_{LX}^\mu(q_1) X^\nu(q_2) \rangle = i F_{LX}^{\mu\nu}(q_1, q_2)~, \qquad &
	\langle h | J_{RX}^\mu(q_1) X^\nu(q_2) \rangle = i F_{RX}^{\mu\nu}(q_1, q_2)~.
	\label{eq:hvJCustFF_B}
\end{split}\ee
Note that, contrary to the previous case, since $e_R$ is a complete singlet of $G$ we can't construct a current $J^{a\mu}_R$ to couple with $L_\mu^a$ or $R_\mu^a$. This is the only difference with respect to the previous case, implying a vanishing $F_{e_R W}^{\mu\nu}$:
\be\begin{split}
	\langle h | J^{a \mu}_{\ell_L} W^{a\nu} \rangle &= i F_{LL}^{\mu\nu}~, \\
	\langle h | J_{e_L}^\mu B^\nu \rangle &= i \left( - \frac{1}{2} c_X F_{LR}^{\mu\nu} + X_{\ell_L} s_X F_{LX}^{\mu\nu} \right) \equiv i F_{e_L B}^{\mu\nu} ~, \\
	\langle h | J_{\nu_L}^\mu B^\nu \rangle &= i \left( \frac{1}{2} c_X F_{LR}^{\mu\nu} + X_{\ell_L} s_X F_{LX}^{\mu\nu} \right) \equiv i F_{\nu_L B}^{\mu\nu} ~, \\
	\langle h | J_{e_R}^\mu W^{3\nu} \rangle &= 0 \equiv i F_{e_R W}^{\mu\nu} ~, \\
	\langle h | J_{e_R}^\mu B^\nu \rangle &= i X_{e_R} s_X F_{RX}^{\mu\nu} \equiv i F_{e_R B}^{\mu\nu} ~.
\end{split}
\label{eq:hvJCustFFSM_B}
\ee
In terms of the physical form factors of Eq.~\eqref{eq:CoeffVJEFT} this setup implies three relations: the two of Eq.~\eqref{eq:CustRelFFVJ} and
\be
	c_w F_{Z e_R}^{\mu\nu} = - s_w F_{\gamma e_R}^{\mu\nu}~.
	\label{eq:CustRelFFVJ_B}
\ee
In the EFT expansion up to $D=6$ terms, Eq.~\eqref{eq:CustRelFFVJ}, this simply becomes
\be
	\epsilon_{Z e_R}^D = 0 ~.
	\label{eq:CustRelCoeffVJ_B}
\ee

\subsubsection*{Summary}

Let us recap the expressions of the contact terms obtained in a custodially-invariant BSM theory.
For example, one can use the first two relations in Eq.~\eqref{eq:CustRelhVVEFT2} to  trade $\epsilon_{W \partial W}$ and $ \epsilon_{Z\partial \gamma}$ for $\delta \kappa_{WZ}$ and $\epsilon_{Z\partial Z}$, and rewrite Eq.~\eqref{eq:contIntCustDecomp} for the two embeddings as
\be\begin{split}
	\epsilon_{W \ell_L^i} &= -\sqrt{2} c_w \epsilon_{Z e_L^i}^D + \frac{g}{\sqrt{2}} \epsilon_{Z\partial Z} + \frac{g}{2 \sqrt{2}} \delta \kappa_{WZ}~,\\
	\epsilon_{Z e_L^i} &= \epsilon_{Z e^i_L}^D - \frac{g}{2c_W} \epsilon_{Z\partial Z} - \frac{g}{2 c_w} \delta\kappa_{WZ}~,\\
	\epsilon_{Z \nu_L^i} &= - \epsilon_{Z e_L^i}^D + \frac{g}{2 c_w} \epsilon_{Z\partial Z}~,\\
	\epsilon_{Z e_R^i} &= \epsilon_{Z e_R^i}^D -  \frac{g}{2 c_w} \delta \kappa_{WZ}~ \quad [\text{embedding } A],\\
	\epsilon_{Z e_R^i} &= -  \frac{g}{2 c_w} \delta \kappa_{WZ}~~~~~ \qquad [\text{embedding } B].
\end{split}\ee
From these expressions, and from Eq.~\eqref{eq:CustRelhVVEFT2}, one easily derives Eqs.~(\ref{eq:custVV}-\ref{eq:custContR}).



\end{document}